\documentclass[pdflatex,sn-mathphys-num]{sn-jnl}% Math and Physical Sciences Numbered Reference Style
\usepackage{array}
\usepackage{makecell}
\newcolumntype{M}[1]{>{\centering\arraybackslash}m{#1}}
\usepackage{graphicx}%
\usepackage{multirow}%
\usepackage{amsmath,amssymb,amsfonts}%
\usepackage{amsthm}%
\usepackage{mathrsfs}%
\usepackage[title]{appendix}%
\usepackage{xcolor}%
\usepackage{textcomp}%
\usepackage{manyfoot}%
\usepackage{algorithmicx}%
\usepackage{listings}%
\usepackage{algorithm}
\usepackage{algpseudocode}
\theoremstyle{thmstyleone}%
\usepackage{float}
\usepackage{booktabs}
\usepackage{subcaption}
\usepackage{siunitx}
\theoremstyle{thmstyletwo}%
\theoremstyle{thmstylethree}%
\usepackage{blkarray}
\usepackage{tabularx}
\usepackage{tikz}
\usepackage{placeins}
\usepackage{booktabs}
\usetikzlibrary{arrows,shapes,quotes}
\usetikzlibrary{patterns,decorations.pathreplacing}

\makeatletter
\newenvironment{breakablealgorithm}
  {% \begin{breakablealgorithm}
   \begin{center}
   \refstepcounter{algorithm}
   \hrule height.8pt depth0pt \kern2pt
   \renewcommand{\caption}[2][\relax]{%
     {\raggedright\textbf{\ALG@name~\thealgorithm} ##2\par}%
     \ifx\relax##1\relax
       \addcontentsline{loa}{algorithm}{\protect\numberline{\thealgorithm}##2}%
     \else
       \addcontentsline{loa}{algorithm}{\protect\numberline{\thealgorithm}##1}%
     \fi
     \kern2pt\hrule\kern2pt
   }}
  {% \end{breakablealgorithm}
   \kern2pt\hrule\relax
   \end{center}
  }
\makeatother

\begin{document}
%\linenumbers
\title[Approximating Parameter Regions for Multistationarity and Multistability in Mass-Action Systems]
{Approximating Parameter Regions for Multistationarity and Multistability in Mass-Action Systems}

\author*[1,2]{\fnm{Dylan Antonio S.J.} \sur{Talabis}}\email{dstalabis1@up.edu.ph}

\author[3,2]{\fnm{Victoria May P.} \sur{Mendoza}}\email{vmpaguio@math.upd.edu.ph}

\author[3,2]{\fnm{Bryan S.} \sur{Hernandez}}\email{bhernandez@math.upd.edu.ph}

\affil*[1]{\orgdiv{Institute of Mathematical Sciences}, \orgname{University of the Philippines Los Ba\~nos}, \orgaddress{\city{Los Ba\~nos}, \country{Philippines}}}

\affil[2]{Data Science Program, College of Science, University of the Philippines Diliman, Quezon City 1101, Philippines}

\affil[3]{\orgdiv{Institute of Mathematics}, \orgname{University of the Philippines Diliman}, \orgaddress{\city{Quezon City}, \country{Philippines}}}

\abstract{Identifying kinetic parameters and conserved quantities that support multistationarity and multistability remains challenging in chemical reaction network analysis. Structural theory provides existence and exclusion results, but does not always yield explicit, interpretable descriptions of the corresponding parameter regimes. We present a computational and machine-learning framework for discovering and characterizing parameter regimes with numerical evidence of multistationarity and multistability in mass-action systems. Here, this numerically detected behavior is referred to as \emph{approximate multistationarity and approximate multistability}. The framework combines Latin hypercube sampling, adaptive sampling, and interpretable surrogate modeling. A logistic-regression model and a neural-network model are separately implemented as alternative surrogates, with each trained on accumulated detection labels and independently used to guide later sampling rounds toward regions with high surrogate scores for approximate multistationarity/multistability. Boundary refinement then targets transition regions between the detected and undetected numerical classes.  Moreover, Buckingham--Pi groups are used in the CN2 rule-induction algorithm to obtain interpretable empirical descriptions of the detected regimes. Illustrative networks yield rules consistent with known multistationarity conditions, and the Schl\"ogl model demonstrates adaptive enrichment of multistationarity detections. In the hybrid histidine kinase model and the sequential dual phosphorylation cycle, the adaptive method substantially increased the number of parameter--total pairs with detected multistability relative to global sampling under equal budgets for sampled parameter. The framework complements analytical chemical reaction network theory methods by providing efficient, parameter-resolved discovery and characterization of multistationary and multistable regimes.}

\keywords{chemical reaction network theory, mass-action kinetics,
	multistationarity, multistability, adaptive sampling,
	interpretable machine learning}

\maketitle

\section{Introduction}\label{sec:intro}

Mass-action kinetics (MAK) provides a mechanistic framework for modeling chemical and biochemical reaction networks. It translates reaction structure into systems of ordinary differential equations whose terms are determined by stoichiometry and reaction rates. Because of this direct correspondence among network structure, stoichiometry, and dynamics, MAK has been widely used to describe many natural systems, particularly well-mixed chemical, cellular, and intracellular processes. Its importance lies not only in its modeling power, but also in its structural transparency: qualitative properties of the resulting dynamical system can often be related to network topology, conservation laws, and kinetic assumptions. This makes MAK especially suitable for the application of rigorous tools from chemical reaction network theory and nonlinear dynamical systems, yielding conclusions that may remain valid across broad ranges of parameter values. \\

Under mass-action kinetics, reaction networks exhibit  intrinsic nonlinear phenomena such as multistationarity, multistability, and oscillations. These behaviors arise from the interplay of feedback mechanisms, conservation laws, stoichiometric constraints, and nonlinear network interactions. In this work, we focus on multistationarity and multistability. A system is multistationary when a stoichiometric compatibility class contains multiple positive equilibria under the same parameter values and conserved quantities. Multistability requires more: at least two of these compatible equilibria, or more generally attractors, must be stable. This distinction is important because the existence of multiple equilibria does not by itself guarantee switch-like behavior; some of the equilibria may be unstable. In biological systems, multistability is particularly significant because it provides a mathematical mechanism for switch-like responses, cellular memory, and robust cell-fate decisions \citep{Harringtonetal2013}. \\

\subsection{Structural Results from Chemical Reaction Network Theory}\label{sec:crnt-structural}
Chemical reaction network theory (CRNT) provides a framework for deriving qualitative conclusions about system dynamics from network structure, often independently of specific parameter values. CRNT establishes rigorous connections between reaction graph structure, stoichiometric constraints, and the behavior of the corresponding mass-action system. One of the most fundamental structural quantities in this theory is the network deficiency. \\

The concept of deficiency was introduced by Feinberg and Horn in the early development of CRNT \citep{Feinberg1972,FeinbergHorn1974}. The deficiency of a chemical reaction network is a nonnegative integer determined by the number of complexes, the number of linkage classes, and the dimension of the stoichiometric subspace. It measures linear dependence between reactions \citep{Shinar2011}. Building on this concept, Feinberg established the Deficiency Zero and Deficiency One Theorems, which provide powerful structural criteria for the existence, uniqueness, and multiplicity of equilibria in mass-action systems \citep{Feinberg1987,Feinberg1995}. \\

The Deficiency Zero Theorem gives one of the clearest structural exclusion results in CRNT. For weakly reversible deficiency-zero networks, it guarantees the existence and uniqueness of a positive equilibrium within each positive stoichiometric compatibility class for all positive rate constants. Consequently, such systems cannot exhibit multistationarity. The Deficiency One Theorem extends uniqueness and multiplicity results to a broader class of networks under more delicate structural assumptions involving connected components. However, these deficiency-based theorems primarily address the existence or exclusion of multiple positive equilibria; they do not, by themselves, generally localize the parameter regions where multistationarity occurs, nor do they automatically establish the stability of the coexisting equilibria. Subsequent work, including extensions by Boros, further developed deficiency-one theory for networks beyond the weakly reversible setting and clarified the existence of positive equilibria in deficiency-one mass-action systems \citep{Boros2011,Boros2013}. \\

Many deficiency-based results on equilibria, particularly those concerning multistationarity, have been made algorithmically accessible through the Chemical Reaction Network Toolbox (CRNToolbox), one of the most widely used software packages in the chemical reaction network theory community \citep{FeinbergEllisonJiKnight2018}. The toolbox implements the deficiency zero and deficiency one theorems, as well as related procedures such as algorithms for networks with higher deficiency. These methods can determine whether a mass-action system has the capacity to admit multiple positive equilibria within the same stoichiometric compatibility class \citep{Ellison1998,Feinberg1972,Feinberg1987,Feinberg1995,Ji2011}. Consequently, they are powerful tools for structural existence and exclusion. However, capacity for multistationarity is an existential statement over parameter space. It does not necessarily identify the parameter values that realize the behavior, and it does not imply stable multistability unless the stability of the resulting equilibria is verified. \\

However, deficiency-based algorithms are not universally applicable. Their validity relies on specific structural assumptions. As a result, there exist reaction networks for which the CRNToolbox algorithms cannot be applied directly. Subsequent works have therefore placed deficiency-based results within a broader framework, combining them with injectivity theory, concordance theory, toric equilibrium methods, and network embedding techniques to assess the capacity for multistationarity in increasingly general classes of reaction networks \citep{JoshiShiu2015}. Complementing these structural approaches, algebraic, geometric, and partitioning methods can sometimes describe multistationarity regions explicitly, particularly in networks with favorable binomial or structured equilibrium forms \citep{ConradiFeliuMinchevaWiuf2017,KaihnsaTelek2024,McClureShiu2024}. These methods represent an important shift from asking whether multistationarity is possible to asking where in parameter space it occurs. \\

\subsection{The Parameter-Space Localization Problem}\label{sec:localization-problem}

Despite the strength of structural CRNT, many results on multistationarity and multistability remain primarily existential. They certify that a network can, or cannot, exhibit a behavior for some admissible parameters, but they often do not indicate where that behavior occurs in the underlying parameter space. This distinction is central: structural capacity concerns the existence of at least one parameter realization, whereas parameter-space approximation seeks to estimate the set of parameter values that are likely to realize the target dynamics. For mass-action systems, this search domain is typically high-dimensional and constrained by positivity, stoichiometric compatibility, and conservation laws. Moreover, multistationary or multistability regimes may occupy thin, disconnected, or geometrically intricate regions. This motivates the \emph{parameter-space approximation problem}: to approximate, identify, and quantify the regions of kinetic parameters and conserved quantities that are likely to exhibit multistationarity or multistability, rather than merely proving that such regions are nonempty. \\

Computational exploration frameworks address this problem by constructing large ensembles of simulations over systematically sampled parameter sets and classifying the resulting trajectories into qualitative dynamical regimes. Instead of analyzing a single nominal parameter vector, these approaches treat parameter space as an object of study in its own right. Candidate parameter sets are generated, the corresponding dynamical systems are simulated from one or more compatible initial conditions, and the resulting time-course behavior is classified according to qualitative features such as convergence to equilibrium, sustained oscillations, chaotic dynamics, multistationarity, or multistability. Computational exploration provides a practical bridge between structural CRNT results, which often establish the possibility of a behavior, and numerical evidence that approximates where such behavior may be realized in parameter space. \\

Latin hypercube sampling and related randomized ensemble strategies are especially useful in this setting because they provide efficient coverage of broad parameter ranges without requiring a full factorial search, which rapidly becomes infeasible in high-dimensional systems. By spreading samples across each parameter dimension, these methods reduce clustering and improve the diversity of explored parameter combinations. When combined with repeated simulations across different initial conditions within the same stoichiometric compatibility class, they allow the parameter space to be empirically partitioned into regions associated with convergence to equilibrium, oscillations, chaos, or multistability \citep{Ferrat2018RF,Qiao2007MAPK,TungLee2010PSExplorer,Brzeski2016BasinStability,Huang2017RACIPE}. These methods are particularly useful for estimating the prevalence and robustness of dynamical regimes under parameter and initial-condition uncertainty. However, because randomized sampling is not guided by the geometry of regime boundaries, it may remain inefficient when the target behavior is rare or localized in a small portion of parameter space. \\

While randomized exploration provides broad coverage, adaptive search methods have been proposed to improve the efficiency of parameter-space investigation. These approaches are motivated by the fact that many dynamical regimes of interest occupy only small, irregular, or highly constrained regions of parameter space, making exhaustive scanning inefficient. Active learning and Bayesian optimization strategies sequentially select informative parameter queries, allowing bifurcations and regime transitions to be located with substantially fewer model evaluations than naive scanning \citep{PsarellisSapsisKevrekidis2025,FaridDR2EI}. By using surrogate models, uncertainty estimates, or acquisition criteria, these methods concentrate simulations near regions where the dynamical classification is uncertain or where transitions between regimes are likely to occur. Related intelligent search strategies based on reinforcement learning have also been used to guide parameter exploration toward rare or complex regimes, such as chaos, by learning policies that reward dynamically informative parameter choices \citep{Huang2025RL}. Adaptive search transforms parameter exploration from a passive sampling problem into an information-driven approximation procedure, making it especially relevant for detecting rare behaviors such as multistability in high-dimensional reaction networks. \\

Machine-learning methods further complement these approaches by providing data-driven tools for classifying qualitative dynamical behavior from large simulation ensembles. Once trajectories have been generated across sampled parameter sets and compatible initial conditions, supervised learning methods can be trained to distinguish between regimes such as convergence to equilibrium, multistationarity, multistability, oscillations, chaos, or other complex attractor structures. Beyond classification, these methods can also help identify the parameters or parameter combinations that most strongly influence regime transitions \citep{CurranHowleyDuggan2025}, recover functional or algebraic relationships associated with multistationarity \citep{Hari2023}, and detect oscillatory or chaotic behavior in biochemical networks \citep{Silk2011}. In this sense, machine learning serves not only as a predictive tool, but also as an exploratory framework for approximating and extracting interpretable patterns from high-dimensional dynamical data. \\

\subsection{Objectives and Main Contributions}
\label{sec:objectives}

In this work, we develop a CRNT-compatible computational framework for
approximating parameter regions with numerical evidence of
multistationarity and multistability in mass-action systems.
The framework searches over kinetic parameters and conserved
quantities, computes equilibrium candidates within stoichiometric
compatibility classes, and distinguishes equilibrium multiplicity
and local stability. Specifically:
\begin{enumerate}
	
	\item We begin with global Latin hypercube sampling of kinetic parameters, together with bounded reference-state sampling and stoichiometrically compatible solver starting points, to establish a baseline numerical search.

	\item We implement adaptive sampling separately using either a class-weighted logistic-regression surrogate or a neural-network surrogate. In each implementation, subsequent rounds combine global exploration with score-guided local proposals and conditional late-stage refinement near the surrogate decision boundary. Here, surrogate outputs strategically guide the search toward promising regions of parameter space.

	\item We use CN2 rule induction with Buckingham--Pi groups to construct
	interpretable empirical descriptions of the detected regimes.
	These rules reveal concise, dimensionless relationships associated with
	multistationarity and multistability across the explored parameter space.
	
	\item We demonstrate the framework through illustrative networks
	with known multistationarity conditions, the Schl\"ogl model,
	and two biochemical case studies: the hybrid histidine kinase
	model and the sequential dual phosphorylation cycle.
\end{enumerate}

The paper is organized as follows. Section~\ref{sec:lhs} introduces basic concepts from chemical reaction network theory and presents the Latin hypercube sampling framework for global exploration of multistationarity, including kinetic-parameter sampling, conservation coordinates, and stoichiometrically compatible starting points. It also introduces Buckingham--Pi groups and CN2 rule induction, with illustrative examples comparing the resulting empirical rules with known analytical multistationarity conditions.

Section~\ref{sec:adaptive} introduces adaptive sampling for rare multistationarity detection, with a class-weighted logistic-regression model and a neural-network model separately implemented as alternative surrogates. Each surrogate is trained on accumulated detection labels and independently used to guide subsequent sampling rounds toward regions with high surrogate scores for approximate multistationarity/multistability. Boundary refinement then targets transition regions between the detected and undetected numerical classes. The Schl\"ogl model illustrates the procedure, round-by-round detection yields, and empirical rules extracted from the sampled data.

Section~\ref{sec:multistability} extends the framework to multistability through local
stability verification within stoichiometric compatibility classes.
Sections~\ref{sec:histidine-kinase} and
\ref{sec:dual-phosphorylation} present the hybrid histidine kinase
and sequential dual phosphorylation case studies, respectively.
These examples compare global and adaptive detection yields and
provide interpretable rule-based descriptions of the evaluated
numerical data.

Section~\ref{sec:conclusion} summarizes the findings, limitations,
and future directions. Detailed equilibrium-search procedures,
CN2 search criteria, surrogate configurations, adaptive-sampling
pseudocode, and restricted-Jacobian stability analysis are provided
in the appendices. 

\section{Global Exploration of Multistationarity}
\label{sec:lhs}

\subsection{Parameters of Chemical Reaction Networks}\label{sec:params}

A chemical reaction network (CRN) is specified by a finite set of reactions among chemical species. Structurally, a CRN can be represented as a directed graph whose vertices are \emph{complexes}, that is, nonnegative linear combinations of species, and whose directed edges are reactions between complexes. Consider the illustrative network below:

\[
\renewcommand{\arraystretch}{0.9}
\begin{array}{c}
	X+Y \xrightarrow{\;k_1\;} 2Y\\[-2pt]
	Y \xrightarrow{\;k_2\;} X
\end{array}
\]

The network is described by the triple $(\mathscr{S},\mathscr{C},\mathscr{R})$ with species set $\mathscr{S}=\{X,Y\}$, complex set $\mathscr{C}=\{X+Y,\,2Y,\,Y,\,X\}$, and reaction set $\mathscr{R}=\{R_1:\,X+Y \xrightarrow{k_1} 2Y,\; R_2:\,Y \xrightarrow{k_2} X\}$, where $k_1,k_2>0$ are the kinetic rate constants. 

For a reaction $R:y\to y'$, the complexes $y$ and $y'$ are called the \emph{reactant} and \emph{product} complexes, respectively. Each reaction $R: y \to y'$ is associated with its \emph{reaction vector} $y'-y$. Hence, the reaction vectors for $R_1$ and $R_2$ are
\[
\begin{array}{l}
	R_1:\; 2Y-(X+Y)=-X+Y=[-1\;\; 1]^\top,\\
	R_2:\; X-Y=[1\;\; -1]^\top.
\end{array}
\]
The \emph{stoichiometric matrix} $N$ is formed by taking these reaction vectors as columns. In this example,
\[
N=\begin{blockarray}{ccc}
	R_1 & R_2\\
	\begin{block}{[cc]c}
		-1 & 1 & X\\
		1 & -1 & Y\\
	\end{block}
\end{blockarray}.
\]

To obtain dynamics, the network is endowed with a kinetics, which assigns a rate function to each reaction. A network $\mathscr N$ together with a kinetics $K$ forms a kinetic system, denoted $(\mathscr N,K)$. Let $[X](t)$ and $[Y](t)$ denote the concentrations of $X$ and $Y$. Under mass-action kinetics, the rate of each reaction is proportional to the product of reactant concentrations raised to their stoichiometric coefficients. For the example above, the rate functions can be written as
\[
\begin{array}{ll}
	R_1: X+Y \to 2Y & \quad K_1 = k_1[X]^1[Y]^1,\\
	R_2: Y \to X & \quad K_2 = k_2[X]^0[Y]^1,
\end{array}
\]
where $k_i>0$ is the rate constant associated with $R_i$. The resulting dynamical system takes the standard form
\[
\frac{dc}{dt}=f(c)=NK(c),
\]
where $c\in\mathbb{R}^{\mathscr S}_{\ge 0}$ collects the species concentrations, $N$ is the stoichiometric matrix, and $K(c)$ is the vector of reaction rates. For this example,
\[
\left[
\begin{array}{c}
	d[X]/dt\\
	d[Y]/dt
\end{array}
\right]
=
\left[
\begin{array}{cc}
	-1 & 1\\
	1 & -1
\end{array}
\right]
\left[
\begin{array}{c}
	k_1[X][Y]\\
	k_2[Y]
\end{array}
\right]
=NK(c),
\]
that is,
\[
\left\{
\begin{aligned}
	\dot{[X]} &= -k_1[X][Y]+k_2[Y],\\
	\dot{[Y]} &= \phantom{-}k_1[X][Y]-k_2[Y].
\end{aligned}
\right.
\]

In this work, the quantities varied in parameter-space exploration naturally fall into two categories: \emph{kinetic rate constants} and \emph{conserved quantities}. The kinetic rate constants \(k=(k_1,\dots,k_r)\in\mathbb{R}^{r}_{>0}\), together with the network structure, determine the reaction-rate vector \(K(c)\), and through the stoichiometric matrix \(N\), uniquely define the vector field \(\dot{c}=NK(c)\) of the mass-action system.

Conservation laws restrict trajectories to affine subsets determined by the stoichiometric matrix. Let $S=\mathrm{im}(N)\subseteq\mathbb{R}^{\mathscr S}$ denote the stoichiometric subspace. For any initial condition $c_0\in\mathbb{R}^{\mathscr S}_{\ge 0}$, solutions remain in the \emph{stoichiometric compatibility class (SCC)}
\[
\mathcal{P}(c_0)=(c_0+S)\cap \mathbb{R}^{\mathscr S}_{\ge 0}.
\]
Equivalently, if $L$ is a matrix whose rows form a basis of the left nullspace of $N$, so that $LN=0$, then $Lc(t)=Lc_0$ along trajectories, and each compatibility class can be written as
\[
\mathcal{P}(c_0)=\{c\in\mathbb{R}^{\mathscr S}_{\ge 0}:\; Lc=Lc_0\}.
\]
These linear equations are known as conservation relations \citep{Dickenstein2016}. In biochemical networks, these invariants often correspond to conserved totals such as enzyme pools or total protein abundances. Consequently, when localizing dynamical regimes, it is often essential to treat conserved totals as parameters on the same footing as rate constants.

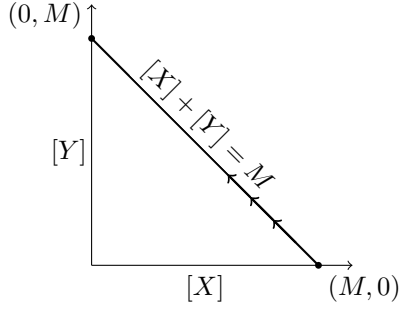
\begin{figure}[t]
	\centering
	\begin{tikzpicture}[scale=3]
		% Axes
		\draw[->] (0,0) -- (1.15,0); 
		\node at (0.5,-0.1) {$[X]$};
		\draw[->] (0,0) -- (0,1.15); 
		\node at (-0.1,0.5) {$[Y]$};
		
		% Stoichiometric class: x+y=1 in the first quadrant
		\draw[thick] (1,0) -- (0,1)
		node[pos=0.55, sloped, above, fill=white, inner sep=2pt]
		{$[X]+[Y]=M$};
		\fill (1,0) circle (0.015) node[below right] {$(M,0)$};
		\fill (0,1) circle (0.015) node[above left] {$(0,M)$};
		
		\draw[->,thick] (0.9,0.1) -- (0.8,0.2);
		\draw[->,thick] (0.8,0.2) -- (0.7,0.3);
		\draw[->,thick] (0.7,0.3) -- (0.6,0.4);
		
	\end{tikzpicture}
	\caption{Stoichiometric compatibility class for the running example. Due to the conservation law of the system, any trajectory originating from 
		a given initial condition is restricted to the invariant line segment $[X]+[Y]=M$ within the nonnegative quadrant.}
	\label{fig:stoich-class-example}
\end{figure}

For the two-species example, the stoichiometric matrix is
\[
N=\begin{bmatrix}-1&1\\[2pt]1&-1\end{bmatrix},
\quad\text{so}\quad
S=\mathrm{im}(N)=\mathrm{span}\{[-1,\,1]^\top\}.
\]
A basis vector for the left nullspace is $w^\top=[1,\,1]$, and thus the scalar quantity $w^\top c=[X]+[Y]$ is conserved. Hence each compatibility class is the line segment in the nonnegative quadrant as illustrated in Figure~\ref{fig:stoich-class-example}.

\subsection{Latin Hypercube Sampling in High Dimensions}\label{sec:lhs-highd}

To explore the vast parameter and state spaces inherent in a kinetic system, we employ Latin Hypercube Sampling (LHS). This method is particularly effective for high-dimensional sensitivity analysis \citep{Hoare2008SaSAT} and existence-focused searches \citep{Tenenbaum2020Robustness}, as it ensures that each marginal distribution is sampled uniformly. \\

The sampling process begins by defining the search hypercube in the parameter space $\Omega_k \subset \mathbb{R}^r_{>0}$ where $r$ is the number of reactions (rate constants). Given the specific bounds $[k_{\min}, k_{\max}]$ for each reaction, the algorithm generates a raw design matrix $U \in [0, 1]^{P_k \times r}$ using a stratified approach  where $P_k$ is the total number of distinct kinetic parameter vectors to be sampled. Hence, for each kinetic rate constant \(k_j\), \(j \in \{1,\dots,r\}\), the interval \([0,1]\) is partitioned into \(P_k\) equally probable bins. A single sample is drawn from each bin such that each row and column of the design matrix contains exactly one sample. The transformation to the physical domain is performed via the function:$$k_{i,j} = k_{min,j} + U_{i,j}(k_{max,j} - k_{min,j})$$ where $k_{i,j}$ represents the $j$-th rate constant for the $i$-th parameter set. This linear scaling ensures that the stratification properties of the LHS are preserved in the target domain. The sampling and labeling procedure is summarized in Algorithm~\ref{alg:global_lhs_scc_label_inlined} in the Appendix.

\subsection{Numerical Computation of Multiple Equilibrium Candidates}
\label{sec:simulation}

Following parameter sampling, equilibrium candidates are computed
by solving the mass-action equilibrium equations directly. For each
sampled pair $(k,c)$, the search uses $n_{\mathrm{IC}}$ starting
points constrained to the corresponding compatibility class and
restricted to the prescribed concentration bounds. This design
distributes a fixed computational budget across many
parameter--compatibility-class pairs while using multiple starts
to search for distinct equilibria within each class.

For fixed rate constants $k$ and conserved quantities $c$, a
mass-action system is \emph{multistationary} on the stoichiometric
compatibility class $\mathcal{P}(c)$ if it admits at least two
distinct positive equilibria:
\[
x^{(1)},x^{(2)}\in\mathcal{P}(c)\cap\mathbb{R}^{m}_{>0},
\qquad
x^{(1)}\neq x^{(2)},
\qquad
f(x^{(1)};k)=f(x^{(2)};k)=0.
\]

To search within a prescribed compatibility class, we represent
states as $x=x_{\mathrm{ref}}+Zz$, where the columns of $Z$ form
an orthonormal basis of the stoichiometric subspace, and solve
$Z^\top f(x_{\mathrm{ref}}+Zz;k)=0$. Without conservation
laws, the full equilibrium equations are solved directly.
Returned candidates are independently checked for finiteness,
mass-action and conservation residuals, and numerical
nonnegativity. Accepted candidates are grouped using a
componentwise absolute--relative clustering criterion to
distinguish repeated recovery of the same candidate from
numerically distinct candidates.

We use \emph{approximate multistationarity} to denote numerical
evidence obtained when at least two distinct candidate
representatives pass these verification tests and the screening criterion. Here,
``approximate'' refers to the tolerance-based verification
of numerical candidates, rather than a relaxation of the
mathematical definition.

Accordingly, a negative detection label means that
multistationarity was \emph{not detected} under the finite
multistart search; it does not establish monostationarity.
The solver settings, verification tolerances, clustering
criterion, boundary screening, and computational details
are provided in Appendix~\ref{app:equilibrium-search}.

\subsection{CN2 Rule Induction}
\label{subsec:positive-rule-mining}

To complement numerical exploration, we use interpretable rules
to describe conditions associated with detected multistationarity.
The procedure follows the CN2
framework \citep{clark1989cn2,clark1991rule} and its subgroup-discovery
adaptation \citep{lavrac2004cn2sd}. 

Let
\[
\mathcal D
=
\{(k_i,c_i,y_i)\}_{i=1}^{n},
\]
where $y_i=1$ denotes detected positive multistationarity according
to the criterion above, and $y_i=0$ denotes that it was not detected.
Labels are inherited from the equilibrium-search output.

From the original and transformed features $\phi(k_i,c_i)$,
the procedure searches directly for rules
\[
A_R(k,c)
=
\bigwedge_{\ell=1}^{L}
\left(
\phi_{j_\ell}(k,c)\circ_\ell\tau_\ell
\right)
\quad\Longrightarrow\quad y=1.
\]
Here, $L$ is the number of conditions, $j_\ell$ identifies a feature,
$\circ_\ell$ is a supported comparison operator, and $\tau_\ell$
is its threshold. Both detected and undetected observations
contribute to evaluation, although the consequent is fixed to
class~1. This distinguishes positive-label enrichment from
conditions merely common in the sampled parameter space.

Rules are induced by beam search and ranked using weighted-relative-
accuracy criteria that balance coverage and positive-label enrichment
\citep{lavrac2004cn2sd}. They are unordered and may describe overlapping
subsets of the sampled parameter space. After each rule is selected,
the weights of covered observations are reduced according to the
configured covering mode, retaining these observations for subsequent
searches while reducing repeated selection. Details of the induction processes, threshold selection, coverage constraints, weight updates, and stopping criteria are provided
in Appendix~\ref{app:cn2-induction}. To evaluate the induced CN2 rules, we compute positive coverage, precision, F1 score, and specificity on the same generated dataset used for rule induction. Here, positive coverage is the fraction of detected cases covered by the rule, precision is the fraction of covered cases that are detected, F1 score balances coverage and precision, and specificity is the fraction of undetected cases not covered by the rule. The corresponding formulas are also provided in the Appendix~\ref{app:cn2-induction}.

\subsubsection{Construction of Buckingham--Pi Transformed Features}
\label{subsec:pi-features}

In our rule induction, we use dimensionless transformed features from rate constants
and selected conserved totals. Dimensional analysis identifies
combinations invariant under changes of physical units
\citep{buckingham1914}. Its integration with data-driven methods
provides a framework for constructing informative dimensionless
representations \citep{bakarji2022,xie2022}. Here, dimensional
validation establishes a feature's physical dimensions; its
association with detected multistationarity is assessed separately
through rule induction.

Assume that all species concentrations share dimension $C$ and
that time has dimension $T$. For reaction $j$, let
\[
\nu_j=\sum_{\ell=1}^{m}y_{\ell j}
\]
be its reactant molecularity. Since $v_j(x;k)$ has dimension
$CT^{-1}$,
\[
[k_j]=C^{1-\nu_j}T^{-1}.
\]
Thus, zero-order, first-order, and second-order rate constants have
dimensions $CT^{-1}$, $T^{-1}$, and $C^{-1}T^{-1}$, respectively.
A conserved pool total formed from concentrations with dimensionless
linear coefficients has dimension $C$.

Let
$\theta=(k_1,\ldots,k_r,c_1,\ldots,c_s)$
collect the scientific variables supplied to feature generation.
Its dimension matrix is
\[
\mathsf D
=
\begin{bmatrix}
	1-\nu_1 & \cdots & 1-\nu_r & 1 & \cdots & 1\\
	-1      & \cdots & -1      & 0 & \cdots & 0
\end{bmatrix}.
\]
Columns contain the concentration and time exponents of the
corresponding variables. Here, the monomial
\[
\Pi_\alpha(\theta)
=
\prod_{j=1}^{r+s}\theta_j^{\alpha_j}
\]
is dimensionless precisely when
\[
\mathsf D\alpha=0.
\]
The implementation enumerates nonzero integer exponent vectors
subject to configured support and exponent bounds:
\[
2\leq\|\alpha\|_0\leq h,
\qquad
|\alpha_j|\leq e,
\]
where $h$ is the maximum support size and $e$ the maximum absolute
exponent.

\subsection{Illustrative Examples}
\label{subsec:illustrative-examples}

Illustrative examples are included to demonstrate the methods discussed above. The illustrative examples are taken from \cite{McClureShiu2024}. These networks provide useful benchmarks because their known multistationarity conditions or qualitative region structures can be compared with the regions detected by the proposed data-driven procedure. For all illustrative examples, the same simulation budget was used. Specifically, the global exploration procedure sampled \(P_k=10{,}000\) distinct kinetic parameter sets using Latin Hypercube Sampling. For each sampled parameter set, \(n_{\mathrm{IC}}=10\) compatible starting points were simulated. Hence, each network was evaluated using a total of \(10{,}000 \times 10 = 100{,}000\) numerical simulations. Table~\ref{tab:network-simulation-bounds} summarizes the reaction networks, conservation structure, and simulation bounds used in the illustrative examples. The first and third networks are one-species systems without
nontrivial conservation laws. The second network is a two-species
conservative system with the conservation law \(A+B=c\). The fourth
network is a two-species system without conservation laws. Across
all examples, rate constants and solver starting points are sampled
within the prescribed lower and upper bounds before numerical
equilibrium computation. For the conservative network, each sampled
reference state determines the conserved total and the corresponding
stoichiometric compatibility class. \\

\begin{table}[htbp]
	\centering
	\small
	\caption{Reaction networks, conservation laws, kinetic parameter
		bounds, and initial-condition bounds used in the global LHS
		exploration.}
	\label{tab:network-simulation-bounds}
	\renewcommand{\arraystretch}{1.20}
	\setlength{\tabcolsep}{3pt}
	
	\begin{tabular}{
			@{}
			p{0.08\textwidth}
			p{0.25\textwidth}
			p{0.22\textwidth}
			p{0.22\textwidth}
			p{0.18\textwidth}
			@{}
		}
		\hline
		&
		\textbf{Reactions} &
		\textbf{Conservation law(s)} &
		\textbf{Rate-constant bounds} &
		\textbf{Starting-point concentration bounds}
		\\
		\hline
		
		\(\mathscr{N}_1\)
		&
		\(\begin{array}{l}
			A \to 0\\
			2A \to 3A\\
			4A \to 3A
		\end{array}\)
		&
		None
		&
		\(\begin{array}{l}
			k_1 \in [0.001,10]\\
			k_2 \in [0.001,10]\\
			k_3 \in [0.001,10]
		\end{array}\)
		&
		\(\begin{array}{l}
			A_0 \in [0.1,100]
		\end{array}\)
		\\
		\hline
		
		\(\mathscr{N}_2\)
		&
		\(\begin{array}{l}
			2A+B \to 3A\\
			A \to B
		\end{array}\)
		&
		\(\begin{array}{l}
			A+B=c
		\end{array}\)
		&
		\(\begin{array}{l}
			k_1 \in [0.001,10]\\
			k_2 \in [0.001,10]
		\end{array}\)
		&
		\(\begin{array}{l}
			A_0 \in [0.1,50]\\
			B_0 \in [0.1,50]
		\end{array}\)
		\\
		\hline
		
		\(\mathscr{N}_3\)
		&
		\(\begin{array}{l}
			A \to 0\\
			0 \to A\\
			2A \to 3A
		\end{array}\)
		&
		None
		&
		\(\begin{array}{l}
			k_1 \in [0.001,10]\\
			k_2 \in [0.001,10]\\
			k_3 \in [0.001,10]
		\end{array}\)
		&
		\(\begin{array}{l}
			A_0 \in [0.1,100]
		\end{array}\)
		\\
		\hline
		
		\(\mathscr{N}_4\)
		&
		\(\begin{array}{l}
			A \to A+B\\
			A+B \to A\\
			2B \to 3B\\
			A \to 2A\\
			2A \to A
		\end{array}\)
		&
		None
		&
		\(\begin{array}{l}
			k_1 \in [0.01,10]\\
			k_2 \in [0.01,10]\\
			k_3 \in [0.01,10]\\
			k_4 \in [0.01,10]\\
			k_5 \in [0.01,10]
		\end{array}\)
		&
		\(\begin{array}{l}
			A_0 \in [0.1,10]\\
			B_0 \in [0.1,10]
		\end{array}\)
		\\
		\hline
	\end{tabular}
\end{table}

Table~\ref{tab:dynamical-classification-results} summarizes the
percentages of parameter sets with detected approximate
multistationarity, the CN2-induced rules, and the corresponding
known multistationarity conditions. Each network was explored
using \(10{,}000\) sampled parameter sets. The learned rules
describe associations with numerical detection rather than
necessary-and-sufficient multistationarity conditions.

The implementation used in the numerical experiments is available
in the GitHub repository
\url{https://github.com/dstalabis1-cmd/Approximating_Parameter_Regions}.
The \texttt{CN2-Pi} module provides dimension-aware feature
construction and CN2 rule induction, while \texttt{Algo\_0}
performs global exploration using Latin Hypercube Sampling
and multistart numerical equilibrium searches.

\begin{table}[htbp]
	\centering
	\small
	\caption{Detected approximate multistationarity percentages,
		CN2-induced rules, and corresponding known multistationarity
		conditions for the four benchmark networks. Percentages are
		computed relative to \(10{,}000\) sampled parameter sets for
		each network.}
	\label{tab:dynamical-classification-results}
	
	\renewcommand{\arraystretch}{1.35}
	\setlength{\tabcolsep}{4pt}
	
	\begin{tabular}{
			@{}
			p{0.12\textwidth}
			p{0.19\textwidth}
			p{0.34\textwidth}
			p{0.31\textwidth}
			@{}
		}
		\hline
		\textbf{Network name}
		&
		\textbf{Detected approximate multistationarity}
		&
		\textbf{CN2-induced rule}
		&
		\textbf{Known multistationarity condition}
		\\
		\hline
		\(\mathscr{N}_1\)
		&
		\(24.72\%\)
		&
		\(\displaystyle
		\frac{k_1^{2}k_3}{k_2^{3}}
		\leq 0.14801\)
		&
		\(\displaystyle
		4k_2^{3}>27k_1^{2}k_3\)
		\\
		\hline
		
		\(\mathscr{N}_2\)
		&
		\(83.86\%\)
		&
		\(\begin{array}{c}
			\displaystyle
			\frac{k_1\,\mathrm{tot}_{A,B}^{\,2}}{k_2}
			\geq 4.2335,\\[2pt]
			\displaystyle
			\mathrm{tot}_{A,B}\leq 69.7958
		\end{array}\)
		&
		\(\displaystyle
		\mathrm{tot}_{A,B}^{\,2}k_1>4k_2
		\)
		\\
		\hline
		
		\(\mathscr{N}_3\)
		&
		\(23.60\%\)
		&
		\(\displaystyle
		\frac{k_1^{2}}{k_2k_3}
		\geq 4.0018\)
		&
		\(\displaystyle
		k_1^{2}>4k_2k_3
		\)
		\\
		\hline
		
		\(\mathscr{N}_4\)
		&
		\(17.96\%\)
		&
		\(\displaystyle
		\frac{k_1k_3k_5}{k_2^{2}k_4}
		\leq 0.2502\)
		&
		\(\displaystyle
		k_2^{2}k_4>4k_1k_3k_5
		\)
		\\
		\hline
	\end{tabular}
\end{table}

The induced rules recover close numerical approximations to the
known multistationarity conditions reported in
\cite{McClureShiu2024}. In particular, the dimensionless features
constructed from the mass-action parameter dimensions allow the
framework to identify combinations of rate constants and conservation
totals that appear in these analytical conditions. 

Although the preceding examples yield
interpretable approximations to known multistationarity conditions,
global sampling may provide relatively few detected cases for
other networks. Consider the four-reaction Schl\"ogl network \citep{Schlogl1972}:
\[
0 \xrightarrow{k_1} X,
\qquad
X \xrightarrow{k_2} 0,
\qquad
2X \xrightarrow{k_3} 3X,
\qquad
3X \xrightarrow{k_4} 2X.
\]
With \(k_j\in[0.001,10]\) for \(j=1,\ldots,4\) and solver
starting-point concentrations in \([0.001,100]\), approximate
multistationarity was detected for only \(183\) of the
\(10{,}000\) sampled parameter sets (\(1.83\%\)). The relatively small
positive set motivates the adaptive sampling stage, which aims to improve the detected positive set, explore regions poorly represented by global sampling, and improve
data-driven descriptions of the target parameter region.

\section{Adaptive Exploration of Multistationarity}
\label{sec:adaptive}

As shown in Section~\ref{sec:lhs}, the effectiveness of global sampling varies
across networks. Rather than relying exclusively on global
sampling, we use numerical detection outcomes from earlier rounds
to guide subsequent sampling toward parameter regions with a
higher predicted probability of detecting multistationarity.
The following subsections describe logistic-regression and
neural-network strategies for enriching the detected positive
set and assess their effectiveness through illustrative examples.

\begin{figure}[h]
	\centering
	\includegraphics[width=\textwidth]{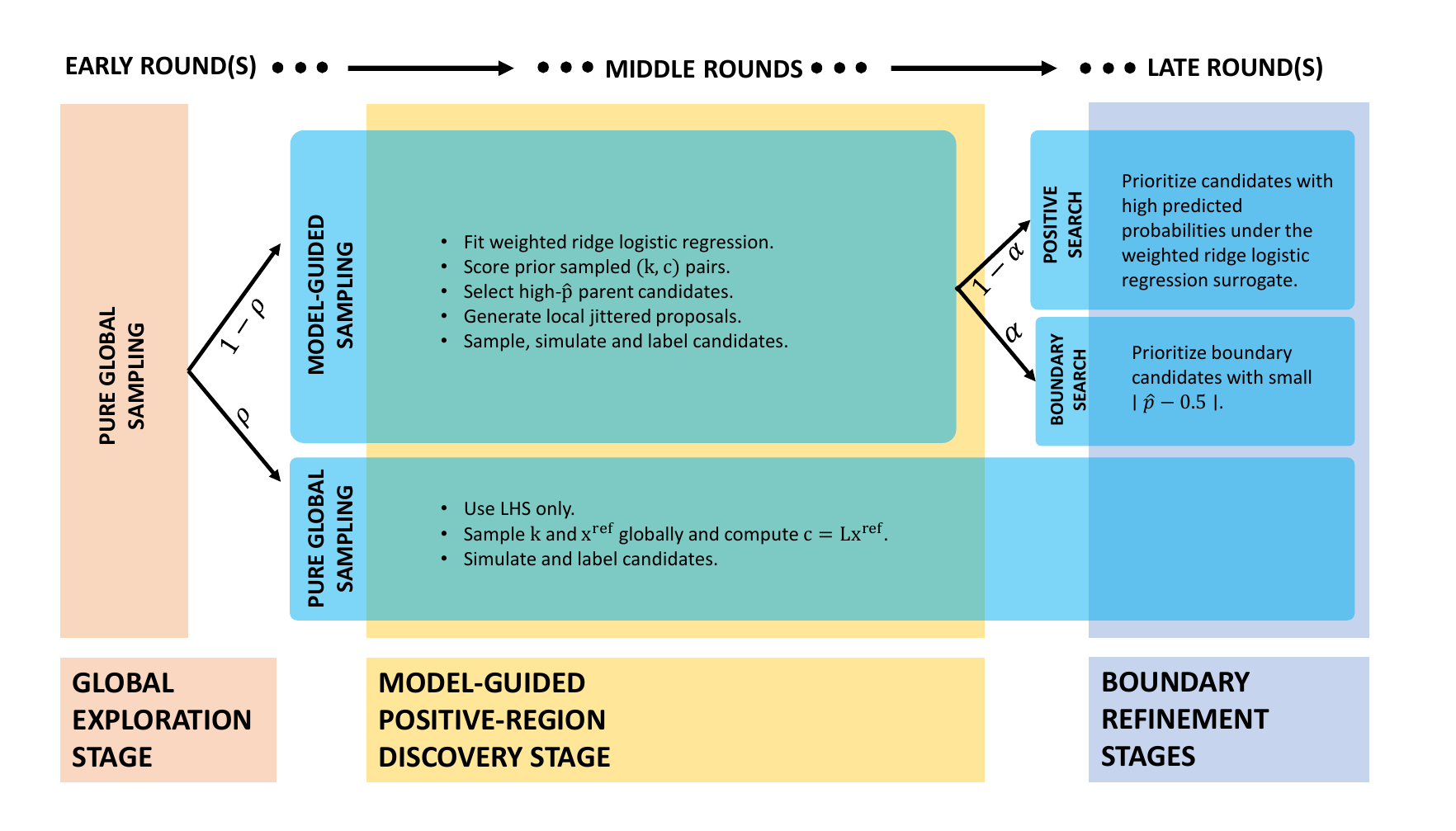}
	\caption{Three-stage structure of the adaptive exploration
		framework. Initial global sampling generates a labeled dataset
		through multistart numerical equilibrium searches. Subsequent
		rounds combine global LHS exploration with local proposals
		ranked by a logistic-regression or neural-network surrogate.
		Conditional late-stage refinement allocates part of the local
		budget to candidates near the surrogate classification boundary.
		Labels describe numerical detection rather than a complete
		characterization of multistationarity.}
	\label{fig:adaptive-overview}
\end{figure}

Figure~\ref{fig:adaptive-overview} summarizes the framework.
An initial global round generates numerical detection labels.
Subsequent rounds combine global exploration with proposals ranked
by a surrogate fitted to the accumulated data. When the specified
late-stage conditions are met, part of the local budget is assigned
to proposals near the surrogate's score-\(0.5\) decision boundary.
Algorithm~\ref{alg:algo1_prob_modified} details the procedure.

\subsection{Reformulation as a Sequential Learning Problem}
\label{sec:sequential}

Recall that a parameter set is represented by \((k,c)\), where
\(k\in\mathbb{R}_{>0}^r\) denotes the vector of rate constants and
\(c=Lx_{\mathrm{ref}}\) denotes the conservation constants defining
the stoichiometric compatibility class (SCC). Here,
\(x_{\mathrm{ref}}\) is an anchor point sampled within the prescribed
concentration bounds. A numerical root-search attempt is specified by
\[
(k,c,x_0),
\]
where \(x_0\) is a solver starting point sampled within the SCC
determined by \(c\). For each sampled pair \((k,c)\), the algorithm searches for
equilibrium candidates from multiple SCC-consistent starting points
and independently verifies and clusters the returned candidates.
Let
\[
q\in\{\mathrm{multistationarity},\mathrm{multistability}\}
\]
denote the selected target, held fixed within an experiment.
The numerical detection label is
\[
Y_{\mathrm{det}}^{(q)}(k,c)=
\begin{cases}
	1, & \text{if the numerical criterion for target \(q\) is met},\\
	0, & \text{otherwise}.
\end{cases}
\]
For multistationarity, the criterion is the detection of at least
two distinct numerically verified interior equilibrium candidates,
as described in Section~2. The multistability criterion additionally
requires stability verification, introduced in Section~4.

The sequential procedure proceeds in rounds \(t=1,2,\dots,T\).
In each round, the algorithm samples or proposes a batch of
\((k,x_{\mathrm{ref}},c)\)-triples, where \(c=Lx_{\mathrm{ref}}\),
and evaluates each pair \((k,c)\) using multiple root-search starts.
Let \(N_t\) denote the cumulative number of evaluated
parameter--compatibility-class pairs after \(t\) rounds. If
\(P_k^{(\tau)}\) pairs are evaluated in round \(\tau\), then
\[
N_t=\sum_{\tau=1}^{t}P_k^{(\tau)}.
\]
For a fixed per-round budget \(P_k\), this reduces to \(N_t=tP_k\).

The accumulated dataset after \(t\) rounds is
\[
\mathcal{D}_t
=
\{(k_i,c_i,y_i)\}_{i=1}^{N_t},
\qquad
y_i=Y_{\mathrm{det}}^{(q)}(k_i,c_i)\in\{0,1\}.
\]
All evaluated pairs enter this history, including those for which
no equilibrium candidate passes verification. The surrogate
predictors are \((k_i,c_i)\), and the response is the pair-level
detection label for the selected target. Before constructing each
subsequent batch, a surrogate is fitted using the accumulated
history, provided both label classes are available and the
learner-specific fitting requirements are satisfied. Otherwise,
the round uses global sampling only. Round~1 is purely global. Rate-constant vectors are generated by
Latin hypercube sampling over the prescribed parameter bounds,
whereas reference states are drawn independently from the uniform
distribution over the reference-state box. In subsequent adaptive
rounds, the global exploration component uses LHS for both the
rate-constant and reference-state coordinates. For a configured
exploration fraction \(\rho\in[0,1]\), the batch is split into
\[
P_{\mathrm{glob}}=\operatorname{round}(\rho P_k),
\qquad
P_{\mathrm{loc}}=P_k-P_{\mathrm{glob}}.
\]
The local component consists of model-guided proposals generated
by perturbing high-scoring historical samples.

Local proposals are generated in \((k,x_{\mathrm{ref}})\)-space
but scored by the surrogate in \((k,c)\)-space. Each proposed
anchor remains within the prescribed concentration bounds and
defines \(c=Lx_{\mathrm{ref}}\). Consequently, its conservation
constants correspond to a nonempty feasible SCC and provide a
reference point for generating SCC-consistent solver starts.
Thus, \(x_{\mathrm{ref}}\) is retained for feasible proposal
generation, while \(c\) is used as a surrogate predictor.

A purely model-driven policy can concentrate on previously
explored regions while overlooking other regions where the target
occurs. Stochastic jitter encourages local neighborhood
exploration, while a nonzero global sampling budget retains
broader exploration. These safeguards limit exclusive reliance
on the current surrogate, but do not guarantee discovery of all
target regions or remove the sampling bias introduced by adaptive
selection.

We consider ridge-regularized logistic regression and a feedforward
neural network as alternative surrogates. Both use standardized
\((k,c)\) predictors and inverse-frequency class weights; the neural
network additionally reserves a stratified validation subset for
training control. Their outputs serve as ranking scores and are not
assumed to be calibrated detection probabilities. Model and training
details are given in Sections~\ref{sec:logreg} and~\ref{sec:neural}
of the appendix.

\subsection{Surrogate-Guided Adaptive Sampling}
\label{sec:prob-guided}

At the start of round \(t\geq2\), we fit a surrogate to the
accumulated dataset \(\mathcal D_{t-1}\), provided its fitting
requirements are satisfied. The resulting score
\(\widehat p_{t-1}(k,c)\) ranks candidates for local sampling.
The logistic-regression and neural-network specifications are
given in Sections~\ref{sec:logreg} and~\ref{sec:neural} of the
appendix. Its output \(\widehat p_{t-1}(k,c)\) is used to rank proposals rather than
assumed to be a calibrated detection probability.

We evaluate \(\widehat p_{t-1}(k_i,c_i)\) on the historical samples
and retain an elite subset \(\mathcal{P}_{t-1}\) containing the
highest-scoring fraction \(\eta\) of previously evaluated pairs,
together with their anchor states. Its size is
\[
|\mathcal{P}_{t-1}|
=
\max\{1,\lceil\eta N_{t-1}\rceil\}.
\]
Parents are drawn stochastically from this elite history using
score-based weights favoring higher-scoring records.
The supplied configuration uses \(\eta=0.20\).

Before selecting local proposals for numerical evaluation, the
algorithm generates an enlarged candidate pool. If
\(P_{\mathrm{loc}}>0\) local proposals are required in round \(t\),
the candidate pool size is
\[
P_{\mathrm{cand}}
=
\max\!\left\{
\left\lceil CF\,P_{\mathrm{loc}}\right\rceil,
5P_{\mathrm{loc}}
\right\},
\]
where \(CF>0\) is the candidate factor.
Thus, the surrogate evaluates more locally perturbed proposals
than will ultimately undergo equilibrium searches.

For each local proposal, let
\[
(k^{(p)},x_{\mathrm{ref}}^{(p)},c^{(p)})
\in\mathcal{P}_{t-1}
\]
denote the selected parent, where
\(c^{(p)}=Lx_{\mathrm{ref}}^{(p)}\).
The superscript \((p)\) identifies the parent.
A candidate pool \(\mathcal{C}_t\) is constructed by perturbing
both its kinetic vector and anchor state:
\[
k^{\mathrm{cand}}
=
R_{\mathcal K}\!\left(k^{(p)}+\varepsilon_k\right),
\qquad
x_{\mathrm{ref}}^{\mathrm{cand}}
=
R_{\mathcal X}\!\left(x_{\mathrm{ref}}^{(p)}+\varepsilon_x\right),
\]
where independent Gaussian perturbations satisfy
\[
\varepsilon_k\sim\mathcal{N}(0,\Sigma_k),
\qquad
\varepsilon_x\sim\mathcal{N}(0,\Sigma_x).
\]
Here, \(R_{\mathcal K}\) and \(R_{\mathcal X}\) denote
componentwise reflection into the kinetic and anchor boxes,
respectively:
\[
\mathcal K=[k_{\min},k_{\max}],
\qquad
\mathcal X=[x_{\min},x_{\max}].
\]
The covariance matrices are diagonal, with scales proportional
to the global coordinate spans:
\[
\Sigma_k
=
\operatorname{diag}\!\left(
(\gamma(k_{\max}-k_{\min}))^2
\right),
\qquad
\Sigma_x
=
\operatorname{diag}\!\left(
(\gamma(x_{\max}-x_{\min}))^2
\right),
\]
where squares are componentwise and \(\gamma\) is the jitter
fraction. The supplied configuration uses \(\gamma=0.15\).
Reflection is used instead of clipping to avoid artificial
accumulation of proposals exactly at the box boundaries.

Each proposed anchor defines
\[
c^{\mathrm{cand}}=Lx_{\mathrm{ref}}^{\mathrm{cand}}.
\]
Thus, although local proposals are generated in
\((k,x_{\mathrm{ref}})\)-space, they are scored in \((k,c)\)-space:
\[
\widehat p_{t-1}\!
\left(k^{\mathrm{cand}},c^{\mathrm{cand}}\right).
\]
When there are no conservation laws, the surrogate uses only \(k\).

Before boundary refinement is activated, the local batch consists
of the \(P_{\mathrm{loc}}\) candidates with the largest surrogate
scores. This concentrates numerical searches on candidates that
the current surrogate ranks as promising for target detection.

In the supplied configuration, boundary refinement is activated
in adaptive rounds satisfying
\[
t\geq\lceil0.80T\rceil,
\qquad
n_{t-1}^{+}\geq40,
\qquad
n_{t-1}^{-}\geq40,
\]
where \(n_{t-1}^{+}\) and \(n_{t-1}^{-}\) count historical
observations with labels \(1\) and \(0\), respectively.
The local budget is then split into
\[
P_{\mathrm{bdry}}
=
\operatorname{round}(0.50P_{\mathrm{loc}}),
\qquad
P_{\mathrm{pos}}
=
P_{\mathrm{loc}}-P_{\mathrm{bdry}}.
\]
The algorithm first selects \(P_{\mathrm{pos}}\) candidates with
the largest scores. From the remaining candidate pool entries,
it selects \(P_{\mathrm{bdry}}\) candidates with the smallest values of
\[
\left|
\widehat p_{t-1}\!
\left(k^{\mathrm{cand}},c^{\mathrm{cand}}\right)-0.5
\right|.
\]
This avoids selecting the same pool entry in both components.
Scores near \(0.5\) identify the current surrogate classification
boundary, not necessarily the actual boundary of the target
parameter region. High-score selection continues alongside
boundary refinement.

The selected local proposals and the \(P_{\mathrm{glob}}\) global
samples form the round-\(t\) batch. Numerical evaluation assigns
each pair its detection label, and the resulting observations
are appended to \(\mathcal D_{t-1}\) to obtain \(\mathcal D_t\).

\subsection{Adaptive Exploration of the Schl\"ogl Network}
To illustrate the adaptive framework, we revisit the Schl\"ogl
network \citep{Schlogl1972}, for which global LHS exploration
detected approximate multistationarity in \(183\) of \(10{,}000\)
sampled parameter sets (\(1.83\%\)). We apply the
logistic-regression and neural-network adaptive strategies with
multistationarity as the selected target. Each adaptive experiment
comprises ten rounds of \(1000\) parameter sets, with ten solver
starting points per set, giving \(10{,}000\) evaluated parameter
sets and \(100{,}000\) root-search attempts.

Table \ref{tab:sampling-stage-counts} reports the detections obtained in each sampling stage. During the model-guided stages, 80\% of the sampling budget is allocated to model-suggested parameter sets. During boundary refinement, the budget is split between model-guided, boundary, and global sampling in proportions of 40\%, 40\%, and 20\%, respectively. Neural-network guidance yielded more detections overall than logistic regression (5,487 (54.87\%) versus 5,426 (54.26\%)), principally through higher model-guided counts in both stages. During boundary refinement, both surrogates yielded 1,200 detections from model-guided sampling. Boundary sampling produced comparable counts, with 597 detections under neural-network guidance and 652 under logistic regression, corresponding to approximately (49.8\%) and (54.3\%), respectively, of the model-guided detection counts during boundary refinement.

\begin{figure}[htbp]
	\centering
	\includegraphics[width=\linewidth]{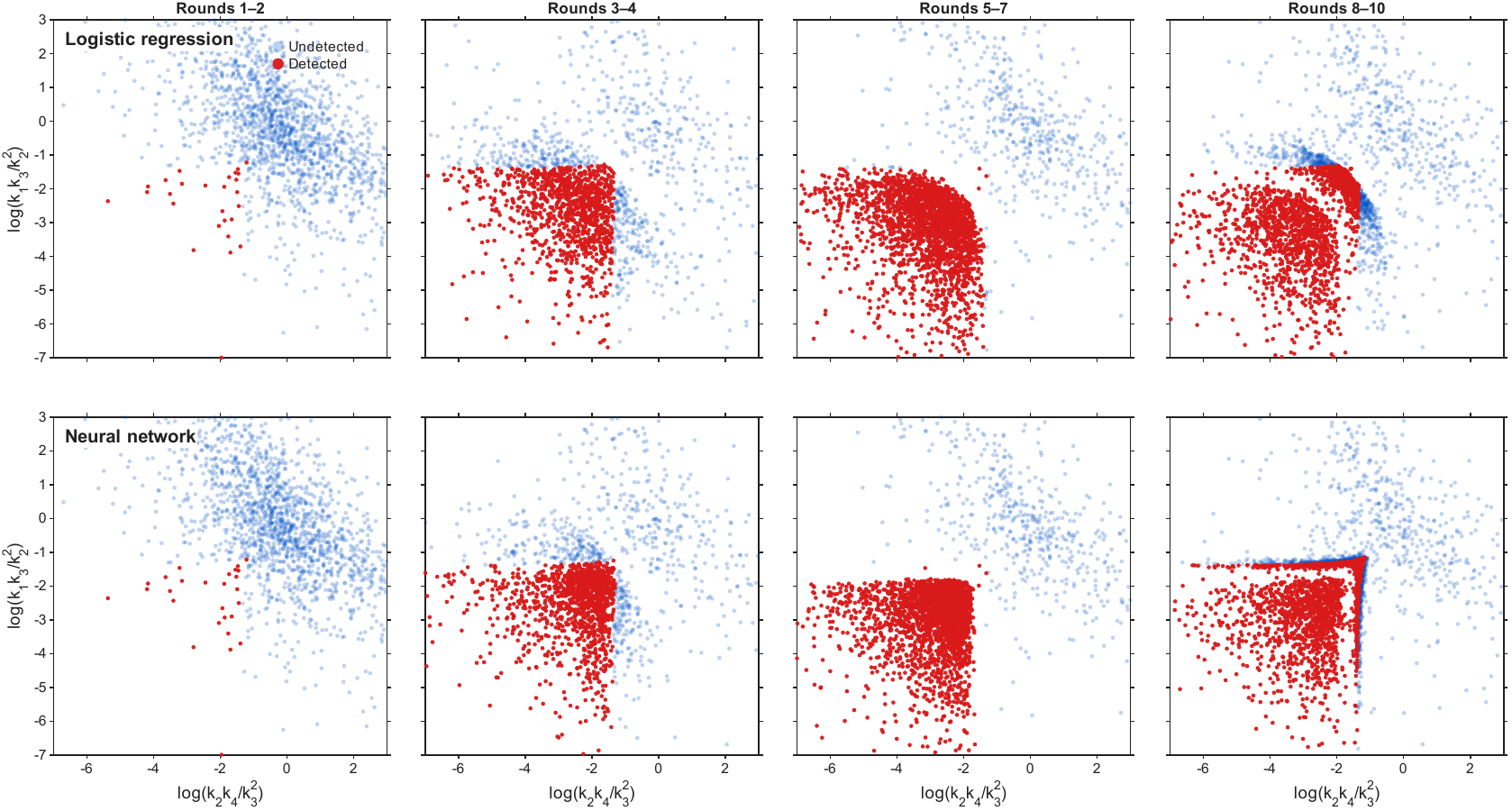}\\[0.5em]
	\includegraphics[width=\linewidth]{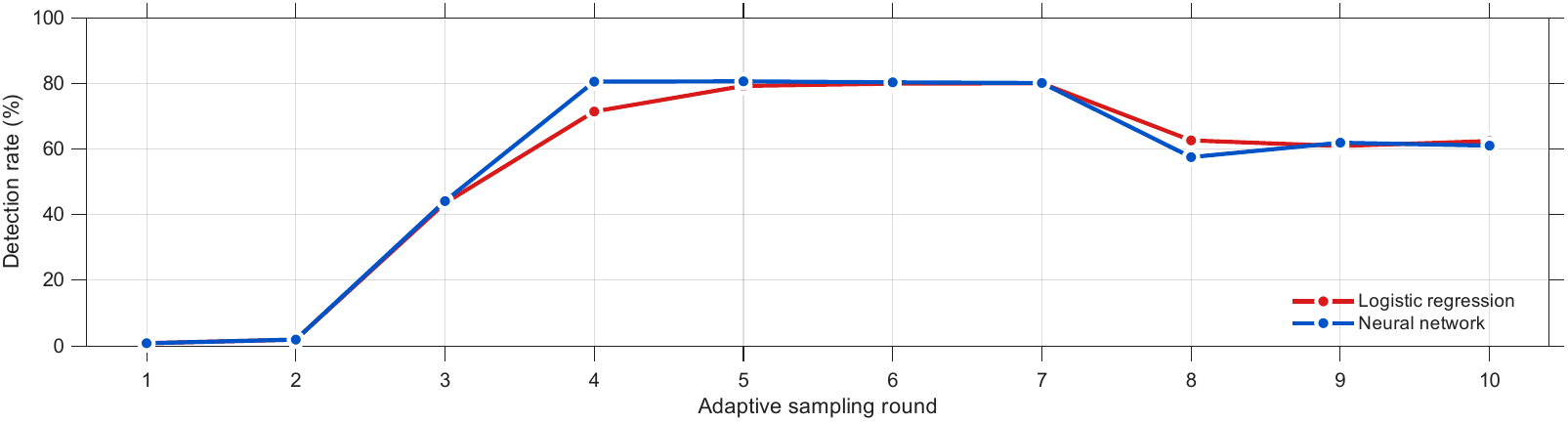}
	\caption{Adaptive sampling for the Schl\"ogl network. The upper and middle rows show sampled parameter sets under logistic-regression and neural-network guidance, respectively, grouped by round. Red points indicate detected approximate multistationarity; blue points indicate no detection. The lower panel shows the detection rate in each round.}
	\label{fig:Excombined}
\end{figure}

The lower part of Figure \ref{fig:Excombined} compares the detection rates across rounds. Neural-network guidance generally achieves a slightly higher rate than logistic regression. Model-guided sampling begins in round 3; the detection rate peaks in round 4 for the neural network and round 5 for logistic regression. Both rates decline modestly during boundary refinement, when sampling shifts toward the edge of the detected region.
The upper part shows the sampled parameter sets by round and stage. The panels show that the adaptive procedure does not sample uniformly after the initial global round. Instead, later rounds increasingly concentrate samples near regions where positive multistationarity parameter sets have already been detected. Along the plotted axes, logistic regression explores a broader, more rounded region, whereas the neural network concentrates samples in region with sharper boundary. This difference is particularly visible during the boundary-refinement stage (rounds 8–10).

\begin{table}[htbp]
	\centering
	\caption{Counts of detected approximate multistationarity by sampling stage and surrogate model.}
	\label{tab:sampling-stage-counts}
	\begin{tabular}{lrrrrrrr}
		\toprule
		& LHS stage
		& \multicolumn{2}{c}{Model-guided stages}
		& \multicolumn{3}{c}{Boundary-refinement stages}
		& Total \\
		\cmidrule(lr){3-4}\cmidrule(lr){5-7}
		Surrogate & Global
		& Model-guided & Global
		& Model-guided & Global & Boundary
		& \\
		\midrule
		Logistic regression & 27 & 3523 & 17 & 1200 & 7 & 652 & 5426 \\
		Neural network      & 27 & 3639 & 17 & 1200 & 7 & 597 & 5487 \\
		\bottomrule
	\end{tabular}
\end{table}

For the Schl\"ogl neural-network dataset, CN2 rule induction produced the following description of the detected class:
\[
\frac{k_2k_4}{k_3^2}\leq 0.31424
\quad\text{and}\quad
\frac{k_1k_3}{k_2^2}\leq 1.57281
\quad\Longrightarrow\quad \mathrm{approx. multistationarity}.
\]
Both antecedent features are dimensionless combinations of the kinetic rate constants. Notably, the first condition is consistent with the theoretically derived necessary condition for multistability of the Schl\"ogl system: 
$\frac{3k_2k_4}{k_3^2}\leq 1$.
Thus, the CN2 threshold \(0.31424\) lies inside the theoretical bound \(1/3\), showing that the data-driven rule recovers a restriction closely aligned with the analytical requirement. On the dataset used to report the rule metrics, positive-class coverage was \(100\%\), meaning that the antecedent covered all parameter sets labeled as detected approximate multistationarity. Precision was \(99.91\%\), meaning that this fraction of the covered observations carried the detected label. The F1 score, balancing precision and positive-class coverage, was \(99.95\%\). Specificity was \(99.82\%\), indicating the fraction of not-detected observations excluded by the antecedent.

\section{Adaptive Exploration of Multistability}
\label{sec:multistability}

Multistability can be difficult to detect through global sampling,
particularly when the target regime is rare within the prescribed
parameter bounds. Here, we extend the framework of Sections~2
and~3 by supplementing equilibrium detection with numerical
stability analysis. For each sampled parameter--compatibility-class
pair $(k,c)$, independently verified equilibrium candidates are
clustered, and boundary or near-boundary representatives are
excluded from the multistability criterion. Approximate
multistability is detected when at least two distinct numerically
interior representatives are classified as locally asymptotically
stable within the same compatibility class.

Stability is assessed using the analytical Jacobian restricted to
the stoichiometric subspace. Candidates are classified as
numerically stable, unstable, or indeterminate according to the
real parts of the restricted eigenvalues and a prescribed
tolerance. The resulting multistability-detection label can replace
the multistationarity label in the adaptive sampling framework.
We apply this procedure to the hybrid histidine kinase model and
the sequential dual phosphorylation cycle. Detailed verification,
clustering, and stability criteria are provided in
Appendix~\ref{app:stability-analysis}.

To obtain interpretable descriptions of the detected multistable region, CN2 rule induction is applied sequentially in two stages. The first stage targets approximate multistationarity. Using the full sampled dataset, the binary response indicates whether a parameter--compatibility-class pair $(k,c)$ satisfies the numerical criterion for approximate multistationarity. CN2 then searches over the available dimensionless and engineered features and produces an antecedent $R_{\mathrm{MS}}$ describing a region enriched in detected multistationary cases. The observations satisfying the selected Stage~1 antecedent are subsequently passed to a second CN2 induction stage, in which the response variable is changed from approximate multistationarity to approximate multistability. Thus, Stage~2 does not attempt to characterize multistability over the entire sampled parameter domain directly. Instead, it refines the region identified in Stage~1 and searches for an additional antecedent $R_{\mathrm{MST}}$ that distinguishes parameter sets with at least two numerically interior locally asymptotically stable equilibria from the remaining observations within the Stage~1 region. The resulting hierarchical rule therefore has the form
\[
R_{\mathrm{MS}}(k,c)
\;\land\;
R_{\mathrm{MST}}(k,c)
\quad\Longrightarrow\quad
\mathrm{multistability}.
\]
This two-stage construction reflects the logical inclusion of multistability within multistationarity: the first stage identifies parameter regions supporting multiple equilibria, while the second stage further resolves whether at least two of those equilibria satisfy the prescribed stability criterion. For interpretability, the conditions obtained from the two stages are retained separately as well as combined into the conjunction $R_{\mathrm{MS}}\land R_{\mathrm{MST}}$. Although the second CN2 model is trained on the subset selected by the Stage~1 rule, the performance of the final combined rule is evaluated with respect to the original dataset. 

\subsection{Case Study I: Hybrid Histidine Kinase Model}
\label{sec:histidine-kinase}

We consider the mass-action hybrid histidine kinase model
studied by  \cite{ConradiFeliuMinchevaWiuf2017}. The kinase has two
phosphorylatable domains, with four states denoted by
$\mathrm{HK}_{00}$, $\mathrm{HK}_{p0}$,
$\mathrm{HK}_{0p}$, and $\mathrm{HK}_{pp}$.
Phosphate transfer to the response regulator $\mathrm{RR}$
produces its phosphorylated form $\mathrm{RR}_{p}$.
The reactions, in the order used to assign the rate constants,
are
\[
\begin{aligned}
	\mathrm{HK}_{00}
	&\xrightarrow{k_1}\mathrm{HK}_{p0},\\
	\mathrm{HK}_{p0}
	&\xrightarrow{k_2}\mathrm{HK}_{0p},\\
	\mathrm{HK}_{0p}
	&\xrightarrow{k_3}\mathrm{HK}_{pp},\\
	\mathrm{HK}_{0p}+\mathrm{RR}
	&\xrightarrow{k_4}\mathrm{HK}_{00}+\mathrm{RR}_{p},\\
	\mathrm{HK}_{pp}+\mathrm{RR}
	&\xrightarrow{k_5}\mathrm{HK}_{p0}+\mathrm{RR}_{p},\\
	\mathrm{RR}_{p}
	&\xrightarrow{k_6}\mathrm{RR}.
\end{aligned}
\]
The conserved totals are
\[
\begin{aligned}
	\mathrm{tot}_{HK}
	&=
	[\mathrm{HK}_{00}]
	+[\mathrm{HK}_{p0}]
	+[\mathrm{HK}_{0p}]
	+[\mathrm{HK}_{pp}],\\
	\mathrm{tot}_{RR}
	&=
	[\mathrm{RR}]+[\mathrm{RR}_{p}],
\end{aligned}
\]
where brackets denote concentrations.
 \cite{ConradiFeliuMinchevaWiuf2017} established that the network
admits multiple positive equilibria for some conserved totals
if and only if $k_3>k_1$. This condition does not guarantee
multistationarity for every choice of totals and does not
itself establish multistability.

For the numerical experiments, all six rate constants were
sampled within $[0.001,10]$. In the species order
\[
(\mathrm{HK}_{00},\mathrm{HK}_{p0},\mathrm{HK}_{0p},
\mathrm{HK}_{pp},\mathrm{RR},\mathrm{RR}_{p}),
\]
the anchor and solver starting-point concentration bounds were
\[
x_{\min}=(1,0,0,0,1,0)^\top,
\qquad
x_{\max}=(50,50,50,50,50,50)^\top.
\]

Global LHS produced only a small number of detections over the prescribed parameter domain, whereas both adaptive strategies substantially increased the detection yield (Table~\ref{tab:multistability-sampling-summary}). The detected fraction increased from $0.27\%$ under global LHS to $18.39\%$ with logistic-regression guidance and $58.91\%$ with neural-network guidance. The improvement was particularly pronounced under neural-network guidance, which produced a considerably larger detected set than logistic-regression guidance in this experiment.

\begin{table}[htbp]
	\centering
	\small
	\caption{Approximate-multistability detection results for the hybrid
		histidine kinase and sequential dual phosphorylation models.
		Each method evaluated $40{,}000$ parameter--compatibility-class
		pairs. Percentages are computed relative to all evaluated pairs.}
	\label{tab:multistability-sampling-summary}
	\renewcommand{\arraystretch}{1.20}
	\setlength{\tabcolsep}{6pt}
	\begin{tabular}{@{}llrr@{}}
		\hline
		\textbf{Model}
		& \textbf{Method}
		& \textbf{Detected pairs}
		& \textbf{Detection (\%)}\\
		\hline
		Hybrid histidine kinase
		& Global LHS   & 108   & 0.27\\
		& Adaptive LR  & 7358  & 18.39\\
		& Adaptive NN  & 23566 & 58.91\\
		\hline
		Sequential dual phosphorylation
		& Global LHS   & 462   & 1.16\\
		& Adaptive LR  & 18355 & 45.89\\
		& Adaptive NN  & 26124 & 65.31\\
		\hline
	\end{tabular}
\end{table}

\begin{figure}[h!]
	\centering
	
	\begin{subfigure}{0.49\textwidth}
		\centering
		\includegraphics[width=\linewidth]{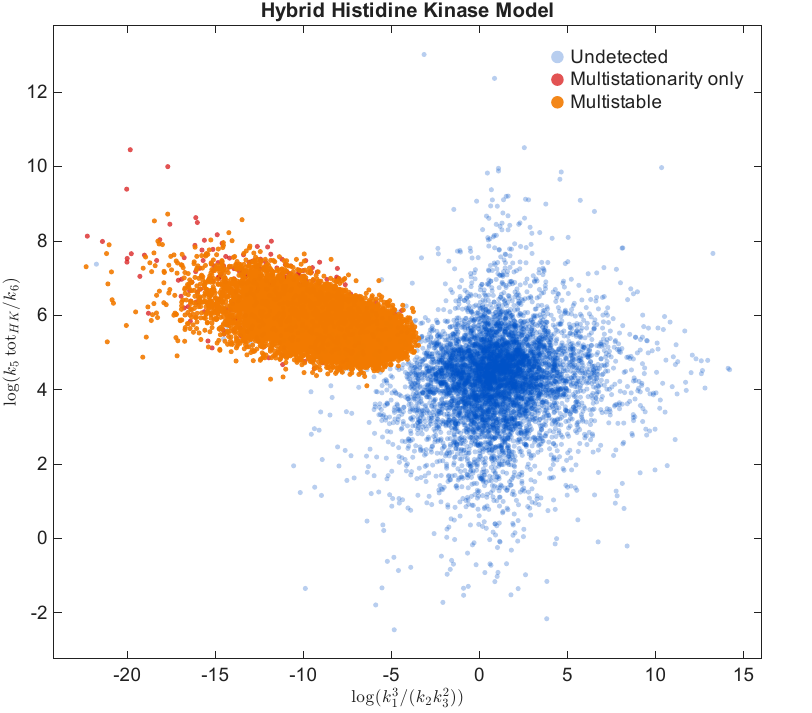}
	\end{subfigure}\hfill
	\begin{subfigure}{0.49\textwidth}
		\centering
		\includegraphics[width=\linewidth]{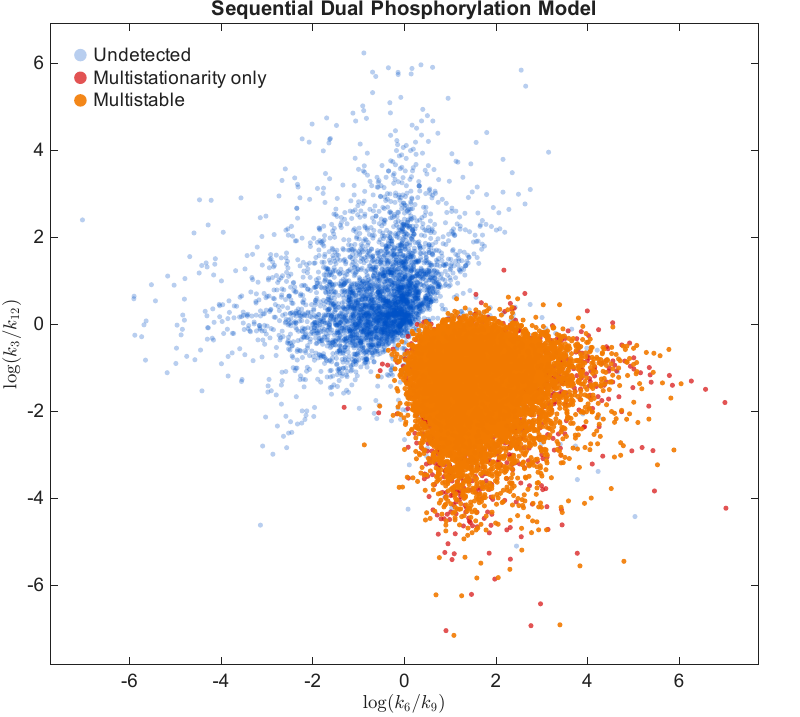}
	\end{subfigure}
	
	\vspace{0.5em}
	
	\includegraphics[width=\textwidth]{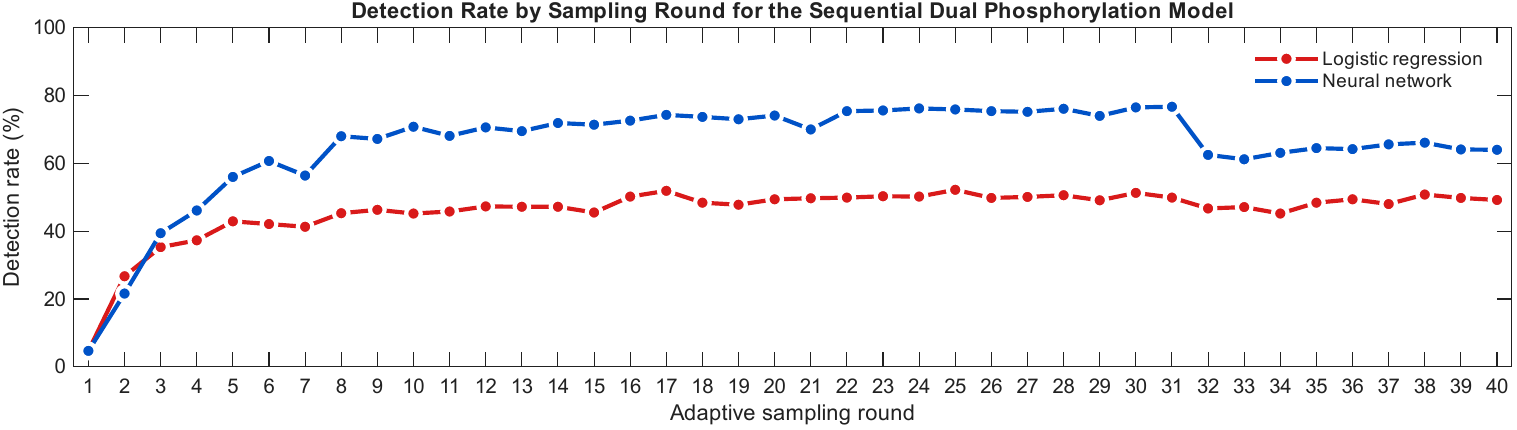}
	
	\vspace{0.5em}
	
	\includegraphics[width=\textwidth]{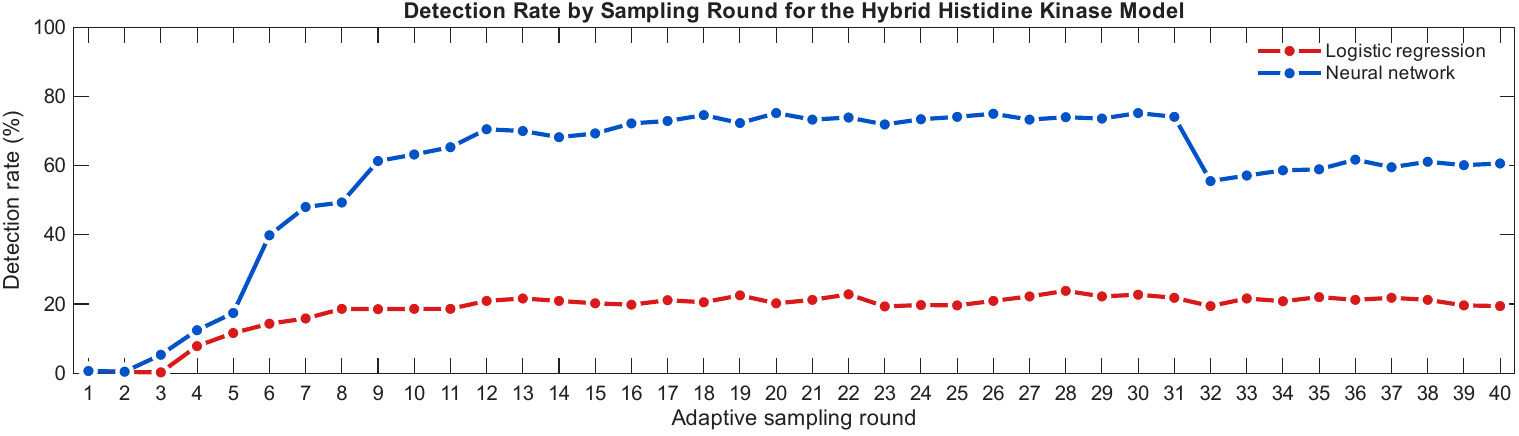}
	
	\caption{Approximate-multistability exploration for the hybrid histidine kinase and sequential dual phosphorylation models. Top panel: projections of the cleaned neural-network dataset onto selected engineered-feature coordinates, with sampled parameter--compatibility-class pairs classified as undetected, approximate multistationarity only, or approximate multistability. Bottom panels: round-wise approximate-multistability detection rates under logistic-regression and neural-network-guided adaptive sampling.}
	\label{fig:combined}
\end{figure}

The lower panel of Figure~\ref{fig:combined} shows the corresponding round-wise detection rates. Detection increases after the initial exploratory rounds as the surrogate begins directing proposals toward regions enriched in approximate multistability. Neural-network guidance exhibits a substantially stronger increase and maintains a higher detection rate through most of the adaptive search, whereas logistic regression reaches a lower but relatively stable detection level. The decrease observed upon entering the later boundary-refinement rounds is consistent with the deliberate reallocation of part of the sampling budget from high-score proposals toward points near the surrogate decision boundary. Thus, the objective of these rounds is not solely to maximize immediate detection yield, but also to improve exploration of the transition between detected and undetected regimes.

To obtain an interpretable description of the detected multistability class, two-stage CN2 rule induction was applied to the neural-network-guided dataset, yielding the rules summarized in Table~\ref{tab:hhk-cn2-rules}. Before rule induction, the dataset was restricted to parameter--compatibility-class pairs for which at least \(90\%\) of the initial-condition root searches produced independently verified accepted equilibria. This filtering reduces the influence of numerical search failures on the learned rules, since insufficiently resolved pairs may be labeled as undetected simply because the equilibrium search failed to recover the relevant equilibria rather than because multistationarity or multistability is genuinely absent. The resulting CN2 analysis therefore emphasizes pairs for which the dynamical classification is supported by comparatively reliable numerical equilibrium searches.

Within this cleaned neural-network dataset, approximately \(97\%\) of the pairs classified as approximately multistationary were also classified as approximately multistable. This high overlap indicates that, for the numerically well-resolved portion of the sampled region, multistationarity was usually accompanied by the stronger stability-based criterion. Accordingly, the rule-performance metrics reported in Table~\ref{tab:hhk-cn2-rules} refer to this cleaned neural-network dataset rather than to the complete adaptive-sampling dataset.

\begin{table}[htbp]
	\centering
	\small
	\caption{Two-stage CN2 rules obtained from the cleaned
		neural-network-guided dataset for the hybrid histidine kinase
		model. Rule-performance metrics are computed on the cleaned
		dataset containing parameter--compatibility-class pairs for which
		at least \(90\%\) of the initial-condition root searches yielded
		independently verified accepted equilibria.}
	\label{tab:hhk-cn2-rules}
	\renewcommand{\arraystretch}{1.15}
	\setlength{\tabcolsep}{5pt}
	\begin{tabular}{@{}cllrrrr@{}}
		\hline
		\textbf{Stage}
		& \textbf{Target}
		& \textbf{Rule}
		& \textbf{Coverage (\%)}
		& \textbf{Precision (\%)}
		& \textbf{F1 (\%)}
		& \textbf{Specificity (\%)}\\
		\hline
		1
		& Multistationarity
		& \(R_1\)
		& 99.672
		& 99.427 
		& 99.549
		& 98.261 \\
		
		1
		& Multistability
		& \(R_1\)
		& 99.664 
		& 96.093
		& 97.846      
		& 89.240 \\
		
		2
		& Multistability
		& \(R_1\land R_2\)
		& 74.48
		& 98.39         
		& 84.781      
		& 96.763 \\
		\hline
	\end{tabular}
	
	\vspace{2mm}
	\begin{minipage}{0.97\linewidth}
		\small
		Here,
		\[
		R_1:\quad
		\frac{k_1^3}{k_2k_3^2}\leq 0.01935,
		\qquad
		\frac{k_5\,\mathrm{tot}_{HK}}{k_6}\geq 88.85765.
		\]
		and
		\[
		R_2:\quad
		\frac{k_5 \mathrm{tot}_{RK}}{k_6}\leq 261.77389,
		\qquad
		\frac{\mathrm{tot}_{HK}}{\mathrm{tot}_{RK}} \leq 3.60173.
		\]
	\end{minipage}
\end{table}

The upper-left panel of Figure~\ref{fig:combined} illustrates the cleaned neural-network-guided dataset projected onto the same engineered features appearing in the CN2 rules. Approximate-multistable pairs are concentrated in a structured region of the projection and are separated from much of the undetected population, indicating that the target regime is associated with a localized geometry in the selected feature space.

\subsection{Case Study II: Sequential Dual Phosphorylation Cycle}
\label{sec:dual-phosphorylation}

The second case study considers the sequential dual phosphorylation
cycle, using an elementary mass-action realization of the second
post-translational modification motif in Table~1 of
\cite{ConradiFeliuMinchevaWiuf2017}. The substrate undergoes two successive
phosphorylation steps catalysed by a kinase, $K$, and two successive
dephosphorylation steps catalysed by a phosphatase, $F$. Enzyme release
after each catalytic step makes the cycle distributive. The network
explicitly includes the four enzyme--substrate complexes:
\begin{equation}
	\begin{aligned}
		A+K
		&\underset{k_2}{\overset{k_1}{\rightleftharpoons}}
		AK \xrightarrow{k_3} A_p+K,\\
		A_p+K
		&\underset{k_5}{\overset{k_4}{\rightleftharpoons}}
		A_pK \xrightarrow{k_6} A_{pp}+K,\\
		A_{pp}+F
		&\underset{k_8}{\overset{k_7}{\rightleftharpoons}}
		A_{pp}F \xrightarrow{k_9} A_p+F,\\
		A_p+F
		&\underset{k_{11}}{\overset{k_{10}}{\rightleftharpoons}}
		A_pF \xrightarrow{k_{12}} A+F.
	\end{aligned}
	\label{eq:dual-phosphorylation-network}
\end{equation}
The nine-species system has three independent conservation coordinates:
\begin{equation}
	\begin{aligned}
		C_1 &= -[A]-[A_p]-[A_{pp}]+[K]+[F],\\
		C_2 &= [K]+[AK]+[A_pK],\\
		C_3 &= [A]+[A_p]+[A_{pp}]-[K]+[A_{pp}F]+[A_pF].
	\end{aligned}
	\label{eq:dual-phosphorylation-conservation-laws}
\end{equation}
Consequently, each stoichiometric compatibility class has dimension six.

For the numerical search, all twelve elementary rate constants were
bounded by $10^{-3}\leq k_j\leq 10$. In the species order
$(A,A_p,A_{pp},K,F,AK,A_pK,A_{pp}F,A_pF)$, the concentration bounds were
\[
x_{\min}
=(1,0,0,0.1,0.1,0,0,0,0)^{\mathsf T},
\qquad
x_{\max}
=(100,100,100,20,20,20,20,20,20)^{\mathsf T}.
\]

We next apply the same framework to the sequential dual phosphorylation cycle. As shown in Table~\ref{tab:multistability-sampling-summary}, both adaptive strategies substantially increased approximate-multistability detection relative to global LHS, with neural-network guidance producing the largest detected set. The detected fraction increased from 1.16\% under global LHS to 45.89\% with logistic-regression guidance and 65.31\% with neural-network guidance, with the latter producing the largest detected set.

Together with the hybrid histidine kinase example, these results show that adaptive sampling can enrich the evaluated dataset in approximate-multistable pairs relative to global LHS. The reported detection rates quantify sampling yield rather than the prevalence of multistability over the original parameter domain.

As in the hybrid histidine kinase case, two-stage CN2 rule induction was applied to the filtered neural-network-guided dataset. Within this cleaned dataset, approximately \(92\%\) of the pairs classified as approximately multistationary were also classified as approximately multistable. The resulting CN2 rules are summarized in Table~\ref{tab:dual-cn2-rules}.

\begin{table}[htbp]
	\centering
	\small
	\caption{Two-stage CN2 rules obtained from the cleaned
		neural-network-guided dataset for the sequential dual
		phosphorylation cycle. Rule-performance metrics are computed
		on the filtered dataset retaining parameter--compatibility-class
		pairs with at least \(90\%\) independently verified accepted
		root-search solutions.}
	\label{tab:dual-cn2-rules}
	\renewcommand{\arraystretch}{1.15}
	\setlength{\tabcolsep}{5pt}
	\begin{tabular}{@{}cllrrrr@{}}
		\hline
		\textbf{Stage}
		& \textbf{Target}
		& \textbf{Rule}
		& \textbf{Coverage (\%)}
		& \textbf{Precision (\%)}
		& \textbf{F1 (\%)}
		& \textbf{Specificity (\%)}\\
		\hline
		1
		& Multistationarity
		& \(R_1\)
		& 98.218 
		& 98.931
		& 98.573 
		& 93.879 \\
		
		1
		& Multistability
		& \(R_1\)
		& 98.367
		& 90.817     
		& 94.441      
		& 64.505 \\  
		
		2
		& Multistability
		& \(R_1\land R_2\)
		& 76.46
		& 94.574
		& 84.558
		& 84.346 \\
		\hline
	\end{tabular}
	
	\vspace{2mm}
	\begin{minipage}{0.97\linewidth}
		\small
		Here,
		\[
		R_1:\quad
		\frac{k_6}{k_9}\geq 1.45331,
		\qquad
		\frac{k_3}{k_{12}}\leq 1.26270.
		\]
		and
		\[
		R_2:\quad
		\frac{C_3}{C_2}\leq 4.57921,
		\qquad
		\frac{k_2k_8^2}{k_3^3}\leq 36.47949.
		\]
	\end{minipage}
\end{table}

The Stage~1 rule \(R_1\) captures nearly all detected approximate-multistationary and approximate-multistable pairs, but its substantially lower specificity for multistability indicates that these two classes are not completely separated by the first pair of features. The Stage~2 condition \(R_2\) increases multistability precision and specificity while reducing positive-class coverage, thereby identifying a more selective region of the detected multistability class. 

\section{Conclusion}
\label{sec:conclusion}

Multistability provides a mechanism for switch-like behavior and memory
in nonlinear chemical and biological systems. However, identifying the
kinetic parameters and conserved totals that support multiple stable
equilibria remains challenging. Chemical reaction network theory
provides powerful tools for establishing the possibility or exclusion
of multistationarity, but theoretical existence results do not always
provide explicit, interpretable descriptions of the parameter regimes
supporting multistability. This motivates a complementary approach to
parameter-resolved numerical discovery.

In this work, we developed a CRNT-compatible computational framework for approximating parameter regions associated with multistationarity and multistability in mass-action systems. Specifically:
\begin{enumerate}
	
	\item We established a global-search baseline using Latin hypercube sampling of kinetic parameters, bounded reference-state sampling, and stoichiometrically compatible solver starts, followed by verification and clustering of equilibrium candidates.
	
	\item We developed adaptive sampling schemes based on class-weighted logistic-regression and neural-network surrogates, combining global exploration, score-guided proposals, and late-stage boundary refinement to improve detection yield.
	
	\item We used CN2 rule induction with dimensionally valid Buckingham--Pi groups and compatible parameter ratios to obtain concise, interpretable descriptions of detected multistationary and multistable regimes.
	
\end{enumerate}

The illustrative multistationarity examples showed that global sampling and rule induction can recover empirical relationships consistent with known analytical conditions, while the Schl\"ogl example demonstrated the benefit of adaptive sampling when positive detections are sparse. The two biochemical case studies extended the same framework to multistability by requiring at least two distinct interior equilibria classified as locally asymptotically stable within the same compatibility class.

Across both biochemical models, adaptive sampling substantially enriched approximate-multistability detections relative to global LHS, with neural-network guidance producing the largest detected sets in the reported experiments. These gains should be interpreted as improved detection yield under the specified sampling and numerical-search procedures

The framework also has important limitations. The framework is necessarily numerical, and its classifications depend on the prescribed parameter bounds, solver initialization, equilibrium-verification tolerances, clustering criteria, and stability thresholds. A finite multi-start search may not recover every equilibrium, particularly in difficult regions of parameter space, and eigenvalues close to the stability threshold can remain numerically ambiguous. Thus, nondetection should be interpreted cautiously rather than as proof of absence. Moreover, the local stability analysis characterizes equilibrium stability within the compatibility class but does not directly resolve attraction basins, switching behavior, or hysteresis.

Adaptive sampling likewise reflects the information accumulated by the surrogate and can preferentially explore already identified positive regions. Surrogate uncertainty should therefore be viewed as a practical guide for refinement rather than as an exact bifurcation boundary. Similarly, CN2 rules provide empirical and interpretable descriptions of numerically detected regimes rather than formal necessary or sufficient conditions. 

Despite these numerical considerations, the results show that the proposed framework can substantially improve the discovery and characterization of rare multistationary and multistable regimes. By integrating compatibility-class equilibrium computation, local stability analysis, adaptive exploration, and dimensionally interpretable rule induction, the framework provides a practical complement to structural and analytical CRNT methods. In particular, it enables efficient exploration of high-dimensional parameter spaces while retaining interpretable descriptions of the regions in which complex dynamical behavior is numerically detected.

\backmatter

\section*{Declarations}
\begin{itemize}
	\item The authors acknowledge the use of artificial intelligence tools, including ChatGPT, to assist in improving the clarity, grammar, organization, and academic style of selected portions of this manuscript. These tools were used only for language refinement, editing support, and formatting suggestions.
	
	\item Funding: This work was supported by the Department of Science and Technology--Science Education Institute (DOST-SEI) through the Accelerated Science and Technology Human Resource Development Program (ASTHRDP) Scholarship Grant.
	\item Conflict of interest/Competing interests: The authors declare no competing interests.
	\item Ethics approval and consent to participate: Not applicable.
	\item Consent for publication: Not applicable.
	\item Data availability: Data are available from the corresponding author upon request.
	%\item Materials availability: Not applicable.
	\item Code availability: The source code and simulation scripts used in this study are available at \url{https://github.com/dstalabis1-cmd/Approximating_Parameter_Regions}. \\
	\item Author contributions: DASJT was responsible for the conceptualization, methodology, validation, and visualization. All authors contributed to the analysis and preparation of the manuscript. BSH and VMPM provided supervision. All authors reviewed and approved the final manuscript.
\end{itemize}

\bibliography{sn-bibliography}% common bib file

\appendix
\section*{Appendix}

\section{Numerical Equilibrium Search and Verification}
\label{app:equilibrium-search}

This section describes the numerical procedure used to compute,
verify, and distinguish equilibrium candidates for each sampled
parameter--compatibility-class pair $(k,c)$. The procedure provides
tolerance-based numerical evidence of equilibrium multiplicity.

\subsection{SCC Selection and Starting-Point Sampling}

For global sampling, each sampled kinetic parameter vector $k_i$
is associated with a reference concentration vector
$x_{\mathrm{ref}}$ drawn uniformly within the concentration bounds
$[x_{\min},x_{\max}]$. This reference state determines the
conservation constants
\[
c_i=Lx_{\mathrm{ref}},
\]
where the rows of $L$ form a basis for the left nullspace of the
stoichiometric matrix $N$. The corresponding stoichiometric
compatibility class is
\[
\mathcal P(x_{\mathrm{ref}})
=
\left\{
x\in\mathbb R_{\geq0}^{m}:
Lx=Lx_{\mathrm{ref}}
\right\}.
\]
Thus, each sampled record consists of a kinetic parameter vector
and a compatibility class, represented by $(k_i,c_i)$.

The conservation constants need not coincide with individual
physical pool totals. Different bases of the left nullspace can
produce signed linear combinations of those totals while defining
the same compatibility class. Consequently, an exported
conservation-coordinate name must be interpreted using its
corresponding row of $L$, including coefficients and signs.

For each sampled pair, multiple positive solver starting points
are generated within the same compatibility class and prescribed
concentration bounds. Let $Z$ have orthonormal columns spanning
$\operatorname{im}(N)$. Candidates are constructed as
\[
x_0=x_{\mathrm{ref}}+Zz
\]
using feasible reduced-coordinate directions and logarithmically
sampled step magnitudes. A hit-and-run--type candidate pool is
also generated, followed by greedy maximin selection in species
log space to diversify the starts. This constrained sampling is
not a standard LHS design.

A candidate starting point is accepted only if it is finite,
strictly positive, within the prescribed concentration bounds,
and consistent with $Lx_0=c_i$ under the sampling tolerance.
Additional feasible samples are generated when necessary;
unfilled positions revert to the reference state. Repeated
reference-state fallbacks do not provide distinct starting points.

Without conservation laws, starting points are sampled directly
in species log space. For species $\ell$, define the effective
positive lower bound
\[
a_\ell
=
\begin{cases}
	x_{\min,\ell}, & x_{\min,\ell}>0,\\
	\max\{10^{-12}x_{\max,\ell},\varepsilon_{\mathrm{sub}}\},
	& x_{\min,\ell}=0,
\end{cases}
\]
where $\varepsilon_{\mathrm{sub}}$ is the smallest positive
subnormal floating-point number. Starting coordinates are generated as
\[
x_{0,\ell}
=
\exp\!\left[
\log a_\ell+
u_\ell(\log x_{\max,\ell}-\log a_\ell)
\right],
\qquad
u_\ell\sim\operatorname{Unif}(0,1).
\]

In the adaptive search, reference states may instead be generated
by global LHS or reflected local proposals, as specified in
Appendix~\ref{app:adaptive-algorithm}. In every case, the
conservation constants are recomputed as $c=Lx_{\mathrm{ref}}$.

\subsection{Reduced Equilibrium Equations}

Let $N\in\mathbb R^{m\times r}$ be the stoichiometric matrix and
$f(x;k)=Nv(x;k)$ the mass-action vector field. Set
$d=\operatorname{rank}(N)$ and let
$Z\in\mathbb R^{m\times d}$ have orthonormal columns spanning
the stoichiometric subspace
\[
S=\operatorname{im}(N)=\ker(L).
\]
In the presence of conservation laws, the equilibrium search uses
\[
x=x_{\mathrm{ref}}+Zz
\]
and solves
\[
G(z;k)=Z^\top f(x_{\mathrm{ref}}+Zz;k)=0.
\]
This representation preserves $Lx=Lx_{\mathrm{ref}}$ in exact
arithmetic. Since $f(x;k)\in S$, the condition
$Z^\top f(x;k)=0$ is equivalent to $f(x;k)=0$.

For an SCC-compatible starting point $x_0$, the initial reduced
coordinate is
\[
z_0=Z^\top(x_0-x_{\mathrm{ref}}).
\]
Without conservation laws, the full system $f(x;k)=0$ is solved.
If $d=0$, the compatibility class is a singleton and the reference
state is checked directly.

\subsection{Nonlinear Solver}

The nonlinear equilibrium equations are solved using
\texttt{fsolve} from MATLAB's Optimization Toolbox
\citep{MATLAB2026b}, with the trust-region dogleg algorithm
and an analytical Jacobian:
\[
J(x;k)=N D_xv(x;k),
\qquad
D_zG(z;k)=Z^\top J(x_{\mathrm{ref}}+Zz;k)Z.
\]
For mass-action kinetics,
\[
v_j(x;k)=k_j\prod_{\ell=1}^{m}x_\ell^{y_{\ell j}}.
\]
For $y_{\ell j}>0$, its derivative is evaluated directly as
\[
\frac{\partial v_j}{\partial x_\ell}
=
k_jy_{\ell j}x_\ell^{y_{\ell j}-1}
\prod_{a\neq\ell}x_a^{y_{aj}},
\]
and is zero when $y_{\ell j}=0$. This avoids division by
$x_\ell$ at zero concentrations.

The residual equations are not divided by a start-dependent
scalar normalization. The supplied configurations allow at most
$300$ iterations and $3000$ function evaluations per solver call,
with function and step tolerances of $10^{-12}$ and an optimality
tolerance of $10^{-10}$.

In the global implementation, Jacobian-based problem scaling is
enabled. If the primary candidate fails independent verification
and the search space has positive dimension, the equations are
retried from the same starting point using the
Levenberg--Marquardt algorithm \citep{MATLAB2026b}. The retry replaces the primary
candidate only if it passes the verification tests below.

The supplied adaptive implementations use the dogleg search
without this Levenberg--Marquardt retry and without explicitly
enabling the global implementation's Jacobian-based problem
scaling. Equal root-start budgets therefore do not imply
identical solver-call budgets or identical solver configurations.

The solver is unconstrained. Intermediate iterates are evaluated
without clipping, and neither positivity nor the starting-point
box is imposed throughout the iterations. Admissibility is
assessed independently after a candidate is returned.

\subsection{Independent Numerical Verification}

A returned candidate $\widehat x$ is retained only if its
coordinates are finite and it passes the full mass-action
residual, conservation residual, and numerical nonnegativity tests.

For every species $\ell$, the mass-action residual must satisfy
\[
|f_\ell(\widehat x;k)|
\leq
\epsilon_{f,\mathrm{abs}}
+
\epsilon_{f,\mathrm{rel}}
\bigl(|N|\,|v(\widehat x;k)|\bigr)_\ell,
\]
where
\[
\epsilon_{f,\mathrm{abs}}=10^{-10},
\qquad
\epsilon_{f,\mathrm{rel}}=10^{-8}.
\]
Matrix and vector absolute values are componentwise. The scaling
term accounts for the gross magnitudes of the reaction
contributions to each species equation. The full mass-action
equations are checked even when the solver uses reduced coordinates.

For every conservation relation $a$, the candidate must satisfy
\[
|(L\widehat x-c)_a|
\leq
\epsilon_{c,\mathrm{abs}}
+
\epsilon_{c,\mathrm{rel}}
\max\!\left\{
|c_a|,(|L|\,|\widehat x|)_a
\right\},
\]
with
\[
\epsilon_{c,\mathrm{abs}}=10^{-10},
\qquad
\epsilon_{c,\mathrm{rel}}=10^{-8}.
\]
This condition is omitted when there are no conservation laws.

Numerical nonnegativity requires
\[
\widehat x_\ell
\geq
-10^{-10}(1+\|\widehat x\|_\infty)
\qquad
\text{for every }\ell.
\]
This test permits small negative coordinates within tolerance;
it does not establish strict positivity. Accepted coordinates
are not clipped before clustering.

Acceptance is determined by these independently recomputed tests,
rather than by the solver exit flag alone. A positive exit flag
is insufficient for acceptance, and a candidate passing the
verification tests may be retained without a positive exit flag.
Consequently, the verified-start count need not equal the count
of positive solver exit flags.

Retained points are called \emph{numerically verified equilibrium candidates}. Verification means satisfaction of the stated
numerical criteria. The concentration box restricts starting points; returned
candidates outside that box may still be accepted if they pass
the verification tests.

\subsection{Clustering of Distinct Candidates}
\label{sec:classification}

Retained candidates are clustered separately for each pair
$(k,c)$ using a greedy componentwise absolute--relative criterion.
A candidate $\widehat x$ is assigned to the first existing
representative $r$ satisfying
\[
|\widehat x_\ell-r_\ell|
\leq
\epsilon_{\mathrm{cl},\mathrm{abs}}
+
\epsilon_{\mathrm{cl},\mathrm{rel}}
\max\{|\widehat x_\ell|,|r_\ell|\}
\qquad
\text{for every }\ell.
\]
The supplied implementations use
\[
(\epsilon_{\mathrm{cl},\mathrm{abs}},
\epsilon_{\mathrm{cl},\mathrm{rel}})
=
\begin{cases}
	(10^{-6},10^{-4}), & \text{global search},\\
	(10^{-8},10^{-5}), & \text{adaptive search}.
\end{cases}
\]
If no representative satisfies the criterion, the candidate becomes
a new representative. Representatives remain retained candidate
points and are not replaced by cluster averages.

Let $\mathcal E(k,c)$ denote the representative set. The count
\[
D_{\mathrm{all}}(k,c)=|\mathcal E(k,c)|
\]
measures distinct numerically verified equilibrium candidates,
including boundary candidates. Multiple accepted starts may
recover one representative, so verified-start counts are
distinguished from distinct-candidate counts.

Clustering defines numerical distinctness at the specified
resolution. Its greedy construction can depend on candidate
order, and sufficiently close candidates may be assigned to
the same representative.

\subsection{Multiplicity, Positivity, and Numerical Labels}

For fixed $(k,c)$, positive multistationarity requires at least
two distinct positive equilibria in the same compatibility
class. Numerical nonnegativity alone is insufficient for
this interpretation.

For each representative $r$, define
\[
\epsilon_{\mathrm b}(r)
=
10^{-10}+10^{-8}\max\{1,\|r\|_\infty\}.
\]
A representative is treated as a numerical boundary or
near-boundary candidate if any coordinate satisfies
$r_\ell\leq\epsilon_{\mathrm b}(r)$. Define
\[
\mathcal E_+(k,c)
=
\left\{
r\in\mathcal E(k,c):
r_\ell>\epsilon_{\mathrm b}(r)
\text{ for every }\ell
\right\},
\qquad
D_+(k,c)=|\mathcal E_+(k,c)|.
\]
This conservative screen may exclude genuinely positive
equilibria close to the boundary.

The implementation's multistationarity flag is
\[
Y_{\mathrm{det}}^{(\mathrm{multistationarity})}(k,c)
=
\mathbf 1\{D_{\mathrm{all}}(k,c)\geq2\}.
\]
The distinct criterion for detected positive multistationarity is
\[
Y_{\mathrm{det},+}^{(\mathrm{multistationarity})}(k,c)
=
\mathbf 1\{D_+(k,c)\geq2\}.
\]
These flags must not be conflated. An exported all-equilibria
multiplicity flag supports a positive-multistationarity
interpretation only when at least two representatives also pass
the interior-state screen. Otherwise, positive-multistationarity
counts require recomputation using $D_+$.

We use \emph{approximate multistationarity} for tolerance-based
numerical evidence of equilibrium multiplicity. The approximation
concerns residual verification and numerical distinctness. A zero label means that the selected target was \emph{not detected} under the finite multistart protocol. Multistability uses the additional
criterion in Appendix~\ref{app:stability-analysis}.

\subsection{Computational Implementation}
\label{sec:parallel-classification}

Parameter vectors and starting points are generated before
parallel execution using deterministic seeds derived from the
base seed and parameter-set index. Root-search attempts are
distributed across workers using MATLAB's \texttt{parfor}.
Attempts within a parameter pair share the same rate constants
and compatibility class.

Outputs retain separate diagnostics for solver termination,
fallback use where applicable, mass-action residuals,
conservation residuals, numerical nonnegativity, and candidate
acceptance. Accepted candidates are associated with
equilibrium-cluster identifiers, while distinct representatives
retain their coordinates, support counts, and boundary status.
Failed attempts remain visible through their diagnostics but
do not enter equilibrium clustering.

Furthermore, generative artificial intelligence tools were used during software development to assist with drafting, debugging, refactoring, and documenting portions of the
MATLAB code. The authors defined the mathematical and
computational methodology, specified the implementation requirements,
reviewed and revised all generated material, executed and tested the
final code, and verified the reported numerical results. No
generative artificial intelligence tool was used to make autonomous
scientific decisions or was credited as an author. The authors take
full responsibility for the code, analyses, interpretations, and final
manuscript.

\subsection{Global Sampling Algorithm}

\begin{algorithm}[H]
	\caption{Global sampling via LHS}
	\label{alg:global_lhs_scc_label_inlined}
	\begin{algorithmic}[1]
		\Require Mass-action model; kinetic bounds
		$k_{\min},k_{\max}\in\mathbb R^r$;
		concentration bounds $x_{\min},x_{\max}\in\mathbb R^m$;
		conservation matrix $L$; orthonormal basis $Z$ of
		$\operatorname{im}(N)$; pair budget $P_k$;
		starts per pair $n_{\mathrm{IC}}$; selected target
		$q\in\{\mathrm{multistationarity},\mathrm{multistability}\}$;
		seed $s$.
		\Ensure Dataset $\mathcal D$ and equilibrium diagnostics.
		
		\State Initialize $\mathcal D\gets\varnothing$ and set RNG to $s$.
		\State Generate a linearly scaled LHS kinetic design
		$K\in[k_{\min},k_{\max}]^{P_k}$.
		\Statex Use randomly permuted stratum midpoints in the primary
		design, or independently randomized positions within strata
		in the fallback design.
		
		\For{$i=1$ to $P_k$}
		\State Retrieve $k_i$ and sample a bounded reference state
		$x_i^{\mathrm{ref}}$ uniformly in the concentration box.
		\State Set $c_i\gets Lx_i^{\mathrm{ref}}$.
		\State Generate $n_{\mathrm{IC}}$ positive, bounded,
		SCC-compatible starting points.
		\State Solve the reduced equilibrium equations from each
		start, or the full equations when there are no conservation laws.
		\State Retain candidates passing independent numerical
		verification and cluster them using the global tolerances.
		\State Record $D_{\mathrm{st}}(k_i,c_i)$ and
		$D_+(k_i,c_i)$.
		\If{$q=\mathrm{multistationarity}$}
		\State Set
		$y_i\gets\mathbf 1\{D_{+}(k_i,c_i)\geq2\}$.
		\Else
		\State Classify interior representatives using the
		restricted Jacobian.
		\State Set
		$y_i\gets\mathbf 1\{D_{\mathrm{st}}(k_i,c_i)\geq2\}$.
		\EndIf
		\State Append $(k_i,c_i,y_i)$ to $\mathcal D$ and retain
		the associated diagnostics.
		\EndFor
	\end{algorithmic}
\end{algorithm}

\section{CN2 Rule Search Algorithm}
\label{app:cn2-induction}

This appendix describes the unordered, positive-target rule
search, threshold construction, weighted covering, and termination
criteria. The supplied implementation supports a standard-WRAcc
mode and an Orange-compatible mode. Their differences are stated
explicitly because they can affect the induced rules. The selected
mode and preprocessing choices must be recorded for each analysis.

\subsection{Inputs and Rule Representation}

Let
\[
\mathcal D=\{(\phi_i,y_i)\}_{i=1}^{n},
\qquad
\phi_i=\phi(k_i,c_i)\in\mathbb R^p,
\]
where $\phi_i$ contains eligible original and transformed features.
The binary response is the numerical detection label selected
for the analysis, which may concern multistationarity or
multistability. Rule induction inherits the supplied labels;
it does not independently verify equilibrium multiplicity
or stability.

Each rule has a fixed positive consequent:
\[
A_R(\phi)
=
\bigwedge_{\ell=1}^{L_R}
(\phi_{j_\ell}\circ_\ell\tau_\ell)
\quad\Longrightarrow\quad y=1.
\]
Its covered-observation set is
\[
C_R=\{i:A_R(\phi_i)\}.
\]
Rules are unordered and may overlap. Their displayed order
records generation order rather than decision-list priority.
The implementation searches directly for class-1 rules and
does not separately optimize class-0 rules.

The configuration specifies the induction mode, beam width $B$,
maximum rule length $L_{\max}$, minimum coverage $m_{\min}$,
and covering factor $0<\gamma_{\mathrm{CN2}}<1$.
Nonfinite feature values never satisfy a condition; observations
are not imputed or silently removed during rule search.

\subsection{Weighted Rule Evaluation}

Observation weights are initialized as $w_i=1$. For a candidate
rule, define
\[
W=\sum_iw_i,
\qquad
W_R=\sum_{i\in C_R}w_i,
\qquad
W_{R,+}=\sum_{i\in C_R}w_i y_i,
\qquad
W_+=\sum_iw_i y_i.
\]
For $W_R>0$, write
\[
p_R=\frac{W_R}{W},
\qquad
p_{+,R}=\frac{W_{R,+}}{W_R},
\qquad
p_+=\frac{W_+}{W}.
\]
Standard weighted relative accuracy is
\[
\operatorname{WRAcc}_w(R)
=
p_R(p_{+,R}-p_+).
\]
It balances weighted coverage against positive-label enrichment
relative to the current weighted dataset
\citep{lavrac2004cn2sd}.

In the standard-WRAcc mode, this expression is used directly.
Only previously unselected antecedents with quality exceeding
a numerical roundoff threshold are selected.

In the Orange-compatible mode, the evaluator is
\[
Q_w(R)
=
\begin{cases}
	p_R(p_{+,R}-p_+), & p_{+,R}>p_+,\\[2pt]
	\displaystyle
	\frac{p_{+,R}-p_+}{\max\{p_R,10^{-6}\}},
	& p_{+,R}\leq p_+.
\end{cases}
\]
The positive branch agrees with standard WRAcc, while the
nonpositive branch penalizes unenriched subsets differently.
This mode does not explicitly require every selected rule
to have positive quality.

Neither mode applies separate statistical-significance or
relative-significance filters. A positive search score therefore
does not constitute a statistical-significance result.

\subsection{Threshold Construction}

Thresholds are constructed using finite feature values of
observations covered by the current parent antecedent. Candidate thresholds are obtained by supervised entropy--MDL
discretization using the current observation weights. For a
feature's distinct sorted values $a_1<\cdots<a_q$, weighted
class counts are aggregated at each value. A candidate split
separates the first $u$ values from the remaining values.

Let $M$ be the total finite-observation weight in the current
discretization subset, and let $H$, $H_{\mathrm L}$, and
$H_{\mathrm R}$ denote the binary class entropies before and
after the split. If $M_{\mathrm L}$ and $M_{\mathrm R}$ are the
child weights, the information gain is
\[
\operatorname{Gain}
=
H-\frac{M_{\mathrm L}}{M}H_{\mathrm L}
-\frac{M_{\mathrm R}}{M}H_{\mathrm R}.
\]
The split maximizing gain is considered first. Let
$K$, $K_{\mathrm L}$, and $K_{\mathrm R}$ be the numbers of
nonzero class counts in the parent and children. Define
\[
\Delta
=
\log_2(3^K-2)
-
\bigl(KH-K_{\mathrm L}H_{\mathrm L}
-K_{\mathrm R}H_{\mathrm R}\bigr).
\]
The split is accepted when $M>1$ and
\[
\operatorname{Gain}
>
\frac{\log_2(M-1)+\Delta}{M}.
\]
Accepted splits are recursively refined in impure children.
If the initial split fails this criterion, the best initial
split is nevertheless retained; this forcing is not applied
to subsequent recursive splits.

The threshold is the first observed feature value on the
right-hand side of a retained split. Conditions use
\[
\phi_j\leq\tau
\qquad\text{or}\qquad
\phi_j\geq\tau.
\]
A refinement must cover positive observations with total weight
at least $m_{\min}$. Except for refinements of the empty
antecedent, its weighted class distribution must differ from
that of its parent. Exact repeated conditions are excluded.

\subsection{Beam Search}

Each covering iteration begins with an empty antecedent covering
all observations. Antecedents are specialized by adding one
condition at a time, up to length $L_{\max}$. At most $B$
candidates remain in the beam after each refinement depth.

Candidate ranking uses quality first and shorter rule length
second. In the standard-WRAcc mode, remaining ties are resolved
by a deterministic canonical condition key, and duplicate
antecedents are removed. In the compatible mode, remaining
ties preserve selector-generation order.

The best eligible rule found during the search is returned,
rather than necessarily the best rule at the final depth.
If no eligible refinement remains, the search stops. The
standard mode returns a new positive-quality antecedent;
the compatible mode returns its highest-scoring valid rule
without an explicit positive-quality restriction.

Beam search is heuristic and does not guarantee a globally
optimal rule set.

\subsection{Weighted Covering and Termination}

After selecting a rule, weights are updated by
\[
w_i^{\mathrm{new}}
=
\begin{cases}
	\gamma_{\mathrm{CN2}}w_i, & i\in H_R,\\
	w_i, & i\notin H_R,
\end{cases}
\]
where
\[
H_R
=
\begin{cases}
	C_R, & \text{standard-WRAcc mode},\\
	C_R\cap\{i:y_i=1\}, & \text{Orange-compatible mode}.
\end{cases}
\]
Thus, the standard mode downweights all covered observations,
whereas the compatible mode downweights only covered positive
observations. Observations remain available for subsequent
searches, permitting partially overlapping rules.

The standard mode stores weights in logarithmic form and
normalizes them during evaluation. Common weight rescaling does
not change standard WRAcc. This mode terminates when no new
positive-quality antecedent is found or when weighted class
masses reach floating-point resolution.

The compatible mode retains absolute observation weights because
minimum coverage depends on weighted positive counts. It
terminates when
\[
\sum_{i:y_i=1}w_i<m_{\min},
\]
when no valid rule is found, or when the best rule covers exactly
the same observation set as an existing rule. No separate maximum
number of rules is imposed.

No rules are induced if there are no eligible features or if
only one label class is present.

\subsection{Reporting and Interpretation}

The induction trace records the weighted positive prior,
covered-observation count, covered-weight fraction, weighted
positive fraction among covered observations, and selection
quality. These training-time quantities are distinguished from
descriptive metrics recomputed on the original CN2 input dataset
with unit weights.

Let $n_+=\sum_i y_i$ and $n_-=n-n_+$. For a rule, define
\[
TP=|C_R\cap\{i:y_i=1\}|,
\qquad
FP=|C_R\cap\{i:y_i=0\}|,
\]
\[
FN=n_+-TP,
\qquad
TN=n_--FP.
\]
The reported metrics, expressed as percentages where appropriate,
are
\[
\operatorname{Coverage}_+=\frac{TP}{n_+},
\qquad
\operatorname{Precision}=\frac{TP}{TP+FP},
\]
\[
F_1=\frac{2TP}{2TP+FP+FN},
\qquad
\operatorname{Specificity}=\frac{TN}{n_-}.
\]

The evaluation population must be stated explicitly. If the
numerical-search history is filtered before CN2 induction,
these metrics describe the filtered dataset, not necessarily
the complete search history.

In the compatible mode, the rule viewer reports the weighted
class distribution at selection and Laplace-corrected percentages.
For weighted class counts $d_0$ and $d_1$, these percentages use
\[
\widehat p_a=\frac{d_a+1}{d_0+d_1+2},
\qquad a\in\{0,1\}.
\]
They are not unsmoothed empirical precision on the original
input observations. The configuration report records the selected mode, covering
factor, feature settings, search limits, and termination reason.
Enumeration and tie handling are deterministic for fixed inputs,
configuration, and row order. Rules exhibiting empirical enrichment describe sampled regions
associated with the selected numerical detection label. 

\section{Surrogate Models for Adaptive Sampling}
\label{app:surrogate-models}

\subsection{Logistic-Regression Surrogate}
\label{sec:logreg}

After each learning round $t$, when both detection classes are
available, a regularized logistic surrogate is fitted to the
accumulated history. Its original-scale representation is
\[
\widehat p_t(k,c)
=
\sigma\!\left(b_0+\beta_k^\top k+\beta_c^\top c\right),
\qquad
\sigma(z)=\frac{1}{1+e^{-z}}.
\]
The response is
$y=Y_{\mathrm{det}}^{(q)}(k,c)\in\{0,1\}$ for the target
$q$ selected in Section~\ref{sec:sequential}. Without conservation
laws, the term involving $c$ is omitted. The output is a sampling
score and is not assumed to be a calibrated detection probability.

Before fitting, predictors are standardized using the accumulated
history:
\[
\widetilde X_{ij}=\frac{X_{ij}-\mu_j}{\sigma_j}.
\]
For constant or numerically invariant features, $\sigma_j$ is
set to one. The ridge penalty acts on standardized slope
coefficients, while the intercept is unpenalized. The supplied
configuration uses $\lambda_{\mathrm{LR}}=10^{-3}$.
Fitted coefficients are subsequently transformed back to the
original feature scale.

Logistic regression is inexpensive to fit and evaluate repeatedly.
Its scores support candidate ranking, while scores near $0.5$
identify candidates near the surrogate decision boundary.
Such candidates need not lie near the actual boundary of the
target parameter region.

The ridge penalty can stabilize slope estimates in sparse
settings \citep{sinkovec2021tune}, while rare outcomes can pose
additional estimation difficulties \citep{king2001logistic}.
Inverse-frequency observation weights balance the two classes:
\[
w_i=
\begin{cases}
	\displaystyle\frac{N_t}{2n_t^+}, & y_i=1,\\[6pt]
	\displaystyle\frac{N_t}{2n_t^-}, & y_i=0,
\end{cases}
\]
where $N_t=|\mathcal D_t|$ and $n_t^+$ and $n_t^-$ are its class
counts. Each class contributes total weight $N_t/2$. This
weighting does not establish probability calibration or correct
all rare-event estimation biases.

The implementation permits up to $100$ Newton iterations with
a relative coefficient-update tolerance of $10^{-8}$. Its
fitting-success flag does not require that this tolerance be
reached: reaching the iteration limit alone does not trigger
global fallback. If only one class is available, or fitting or
historical scoring is unavailable or unsuccessful, the next
batch uses global sampling only.

\subsection{Neural-Network Surrogate}
\label{sec:neural}

The alternative surrogate is a feedforward neural network fitted
to the accumulated detection history. Predictors remain the
rate constants and conservation coordinates $(k,c)$, and the
response is the selected pair-level detection label.

Let
\[
u=\begin{pmatrix}k\\c\end{pmatrix}
\]
and let $\widetilde u$ be its standardized representation,
with standardization fitted using the training subset.
Without conservation laws, $u=k$.

The network has two fully connected hidden layers with $64$
and $32$ units:
\[
h_1=\phi(W_1\widetilde u+b_1),
\qquad
h_2=\phi(W_2h_1+b_2),
\]
where $\phi(z)=\max\{0,z\}$ acts componentwise.
A final fully connected layer produces two logits:
\[
a=W_3h_2+b_3,
\qquad a=(a_0,a_1)^\top.
\]
The sampling score is the detected-class softmax output:
\[
\widehat p_t(k,c)
=
\frac{\exp(a_1)}{\exp(a_0)+\exp(a_1)}.
\]

Inverse-frequency weights are computed from the accumulated
history before its training--validation split. MATLAB's
\texttt{fitcnet} fits the network using weighted cross-entropy
with ridge regularization through an L-BFGS optimizer.
The supplied configuration uses
$\lambda_{\mathrm{NN}}=10^{-4}$, He initialization for layer
weights, and zero initialization for biases.

A seeded stratified split reserves approximately $20\%$ of
each class for validation and retains both classes in both
subsets. Fitting requires at least $10$ historical observations
of each class. Training is limited to $300$ optimizer iterations,
with validation loss evaluated every iteration. Early stopping
uses patience $10$: training stops after ten consecutive
validation checks without a new minimum validation loss.

The validation subset controls training; it is not an independent
test over the global parameter domain. A fitted network is used
only when its training and validation loss diagnostics are
finite and its scores on the accumulated history are finite
and numerically nonconstant. Otherwise, the next batch uses
global sampling only.

As with weighted logistic regression, scores are used for ranking
rather than assumed to be calibrated probabilities. Numerical
equilibrium search and stability verification, not surrogate
predictions, determine new detection labels.

\section{Complete Algorithm for Adaptive Sampling}
\label{app:adaptive-algorithm}

The adaptive procedure retains global exploration while using
surrogate scores to rank locally proposed parameter--reference-state
pairs. Local proposals are formed in the original linear
coordinates, not in logarithmic parameter coordinates.
Conservation coordinates are recomputed from proposed reference
states rather than perturbed independently.

The elite pool contains
\[
P_{\mathrm{elite}}
=
\max\{1,\lceil\eta N_{t-1}\rceil\}
\]
highest-scoring historical records. Parent-selection weights are
proportional to their scores in the LR implementation; the NN
implementation applies a small positive floor before normalization.

\begin{breakablealgorithm}
	\caption{Surrogate-guided adaptive sampling}
	\label{alg:algo1_prob_modified}
	\begin{algorithmic}[1]
		\Require Mass-action model; kinetic bounds
		$k_{\min},k_{\max}\in\mathbb R^r$;
		concentration bounds $x_{\min},x_{\max}\in\mathbb R^m$;
		conservation matrix $L$; orthonormal basis $Z$;
		target $q\in\{\mathrm{multistationarity},\mathrm{multistability}\}$;
		learner $M\in\{\mathrm{LR},\mathrm{NN}\}$;
		rounds $T$; pairs per round $P_k$; starts per pair
		$n_{\mathrm{IC}}$; exploration fraction $\rho$;
		elite fraction $\eta$; candidate factor $CF$;
		jitter fraction $\gamma_{\mathrm{jit}}$; seed $s$.
		\Ensure Indexed detection history $\mathcal D_T$ and
		parameter--reference-state history $\mathcal H_T$.
		
		\State Define $\mathcal K=[k_{\min},k_{\max}]$ and
		$\mathcal X=[x_{\min},x_{\max}]$.
		\State Define $\Delta_k=k_{\max}-k_{\min}$ and
		$\Delta_x=x_{\max}-x_{\min}$.
		\State Set
		\[
		\Sigma_k=\operatorname{diag}((\gamma_{\mathrm{jit}}\Delta_k)^2),
		\qquad
		\Sigma_x=\operatorname{diag}((\gamma_{\mathrm{jit}}\Delta_x)^2).
		\]
		\State Set $\alpha=0.50$ and
		$t_{\mathrm b}=\lceil0.80T\rceil$.$^*$
		\State Initialize $\mathcal D_0=\mathcal H_0=\varnothing$.
		
		\For{$t=1$ to $T$}
		\State Set $s_t=s+1000(t-1)$.
		\State Initialize $\mathcal D_t\gets\mathcal D_{t-1}$
		and $\mathcal H_t\gets\mathcal H_{t-1}$.
		
		\If{$t=1$ or fewer than two historical label classes}
		\State Generate a full global kinetic LHS batch
		using the base-seed branch.
		\State Generate bounded uniform reference states
		using the implementation's deterministic anchor seeds.
		\Else
		\State Attempt class-weighted fitting and historical
		scoring with learner $M$.
		\If{fitting or historical scoring is unavailable
			or unsuccessful}
		\State Set RNG to $s_t$ and generate full
		global LHS batches in $\mathcal K$ and $\mathcal X$.
		\Else
		\State Let $\widehat p_{t-1}(k,c)$ be the fitted score.
		\State Select
		$\max\{1,\lceil\eta N_{t-1}\rceil\}$
		highest-scoring historical records as the elite pool.
		\State Construct normalized learner-specific
		score-based parent weights.
		\State Set
		\[
		P_{\mathrm{glob}}=\operatorname{round}(\rho P_k),
		\qquad
		P_{\mathrm{loc}}=P_k-P_{\mathrm{glob}}.
		\]
		\State Set RNG to $s_t$ and generate
		$P_{\mathrm{glob}}$ global LHS records in
		$\mathcal K$ and $\mathcal X$.
		\State Initialize empty local sampling arrays.
		
		\If{$P_{\mathrm{loc}}>0$}
		\State Set
		\[
		P_{\mathrm{cand}}
		=
		\max\{\lceil CF P_{\mathrm{loc}}\rceil,
		5P_{\mathrm{loc}}\}.
		\]
		\State Set RNG to $s_t+1$.
		
		\For{$b=1$ to $P_{\mathrm{cand}}$}
		\State Sample a parent
		$(k^{(p)},x_{\mathrm{ref}}^{(p)},c^{(p)})$
		from the elite pool using its normalized weights.
		\State Independently sample
		$\varepsilon_k\sim\mathcal N(0,\Sigma_k)$
		and
		$\varepsilon_x\sim\mathcal N(0,\Sigma_x)$.
		\State Form
		\[
		k_b^{\mathrm{cand}}
		=
		R_{\mathcal K}(k^{(p)}+\varepsilon_k),
		\qquad
		x_b^{\mathrm{cand}}
		=
		R_{\mathcal X}
		(x_{\mathrm{ref}}^{(p)}+\varepsilon_x).
		\]
		\State Set
		$c_b^{\mathrm{cand}}=Lx_b^{\mathrm{cand}}$.
		\State Compute
		$p_b^{\mathrm{cand}}
		=
		\widehat p_{t-1}
		(k_b^{\mathrm{cand}},c_b^{\mathrm{cand}})$.
		\EndFor
		
		\If{$t\geq t_{\mathrm b}$ and both historical
			class counts are at least $40$}
		\State Set
		$P_{\mathrm{bdry}}
		=\operatorname{round}(\alpha P_{\mathrm{loc}})$
		and
		$P_{\mathrm{pos}}=P_{\mathrm{loc}}-P_{\mathrm{bdry}}$.
		\State Select $P_{\mathrm{pos}}$ candidates
		with largest scores.
		\State Select $P_{\mathrm{bdry}}$ remaining
		candidates with smallest
		$|p_b^{\mathrm{cand}}-0.5|$.
		\Else
		\State Select $P_{\mathrm{loc}}$ candidates
		with largest scores.
		\EndIf
		\State Extract the selected kinetic and
		reference-state rows.
		\EndIf
		\State Concatenate global and local records,
		preserving their row correspondence.
		\EndIf
		\EndIf
		
		\For{$b=1$ to $P_k$}
		\State Set global record index $i=(t-1)P_k+b$.
		\State Retrieve corresponding $k_i$ and
		$x_i^{\mathrm{ref}}$; set $c_i=Lx_i^{\mathrm{ref}}$.
		\State Generate $n_{\mathrm{IC}}$ positive, bounded,
		SCC-compatible starting points.
		\State Solve the reduced or full equilibrium equations
		using the adaptive solver configuration.
		\State Independently verify candidates and cluster them
		using the adaptive tolerances.
		\If{$q=\mathrm{multistationarity}$}
		\State Set
		$y_i=\mathbf 1\{D_{\mathrm{+}}(k_i,c_i)\geq2\}$.
		\Else
		\State Classify interior representatives using
		restricted-Jacobian eigenvalues.
		\State Set
		$y_i=\mathbf 1\{D_{\mathrm{st}}(k_i,c_i)\geq2\}$.
		\EndIf
		\State Append $(k_i,c_i,y_i)$ to $\mathcal D_t$
		and $(k_i,x_i^{\mathrm{ref}},c_i)$ to $\mathcal H_t$.
		\EndFor
		\EndFor
		
		\Statex \textbf{Notation and implementation notes.}
		$R_{\mathcal K}$ and $R_{\mathcal X}$ perform componentwise
		reflection into their respective boxes. Covariance squares are
		componentwise. Empty sampling blocks contain zero rows.
		Historical records are indexed observations rather than
		deduplicated parameter sets; repeated records are retained.
		Without conservation laws, $c$ is omitted.
		The one-class global branch can reuse the base kinetic LHS
		design and should not be interpreted as a fresh independent
		design in every round. $^*$Boundary refinement is activated only when
		$t\geq\lceil0.80T\rceil$ and at least $40$ observations of each
		detection class are available. Half of the local allocation is
		then selected by high score and half by proximity to $0.5$,
		using distinct candidate-pool indices.
	\end{algorithmic}
\end{breakablealgorithm}

\section{Stability Verification and Restricted-Jacobian Analysis}
\label{app:stability-analysis}

\subsection{Stability Verification of Detected Equilibria}
\label{sec:stability-verification}

Sections~2 and~3 concern equilibrium multiplicity without requiring
stability. For fixed $(k,c)$, positive multistability means that the
compatibility class contains at least two distinct positive
equilibria that are locally asymptotically stable relative to that
class. Multiple equilibria alone are therefore insufficient.

The search is root-based: starting points initialize a nonlinear
solver rather than numerically integrated trajectories. Returned
candidates undergo the independent verification and clustering
procedure in Appendix~\ref{app:equilibrium-search}. The appropriate
global or adaptive clustering tolerances are used.

Stability is recomputed at each retained representative, not at
a cluster average. Repeated recovery of one representative does
not increase the distinct-equilibrium count.

Only representatives in
\[
\mathcal E_+(k,c)
=
\left\{
r\in\mathcal E(k,c):
r_\ell>\epsilon_{\mathrm b}(r)
\text{ for every }\ell
\right\}
\]
are classified by the interior-state stability test.
Boundary and near-boundary candidates are flagged separately,
left unclassified by this test, and excluded from multistability
detection.

Let
$\mathcal E_{\mathrm{st}}(k,c)\subseteq\mathcal E_+(k,c)$
contain representatives classified as numerically stable by the
criterion below. Define
\[
D_{\mathrm{st}}(k,c)=|\mathcal E_{\mathrm{st}}(k,c)|,
\]
\[
Y_{\mathrm{det}}^{(\mathrm{multistability})}(k,c)
=
\mathbf 1\{D_{\mathrm{st}}(k,c)\geq2\}.
\]
A positive label denotes detected approximate multistability under
the numerical protocol, not rigorous certification of exact
equilibria and their stability. A zero label means not detected,
not certified monostability. 

\subsection{Restricted-Jacobian Eigenvalue Analysis}
\label{sec:restricted-jacobian}

For $f(x;k)=Nv(x;k)$, the analytical Jacobian is
\[
J(x;k)=N D_xv(x;k).
\]
Let $d=\operatorname{rank}(N)$ and let
$Z\in\mathbb R^{m\times d}$ have orthonormal columns spanning
$S=\operatorname{im}(N)$. At a retained representative $r$,
the restricted Jacobian is
\[
J_S(r;k)=Z^\top J(r;k)Z.
\]
Because $J$ maps perturbations into $S$, this matrix represents
the linearized dynamics of perturbations within the compatibility
class. It excludes the structural zero eigenvalues associated
with conservation laws without removing genuine zero eigenvalues
of the within-class dynamics. Without conservation laws,
$Z=I_m$ and $J_S=J$.

For $d>0$, define the computed spectral abscissa
\[
\alpha(r;k)
=
\max_{\lambda\in\operatorname{spec}(J_S(r;k))}
\operatorname{Re}(\lambda)
\]
and the stability tolerance
\[
\tau_{\mathrm{st}}(r;k)
=
10^{-10}
+
10^{-7}\max\{1,\|J_S(r;k)\|_2\},
\]
where $\|\cdot\|_2$ is the matrix spectral norm.
Each numerically interior representative is classified as
\[
\operatorname{Class}(r;k)
=
\begin{cases}
	\text{numerically stable},
	& \alpha(r;k)<-\tau_{\mathrm{st}}(r;k),\\[2pt]
	\text{numerically unstable},
	& \alpha(r;k)>\tau_{\mathrm{st}}(r;k),\\[2pt]
	\text{indeterminate},
	& |\alpha(r;k)|\leq\tau_{\mathrm{st}}(r;k).
\end{cases}
\]

At an exact positive equilibrium, strictly negative real parts
of all restricted eigenvalues imply local asymptotic stability
relative to the compatibility class. A restricted eigenvalue with
positive real part implies instability. The numerical tolerance
introduces an indeterminate band around the imaginary axis;
membership in that band does not establish neutral stability
or a bifurcation.

Outputs record the spectral abscissa, stability tolerance,
and numbers of eigenvalues with real parts below
$-\tau_{\mathrm{st}}$, above $\tau_{\mathrm{st}}$, or within
the tolerance band. Boundary candidates are flagged separately
and are not subjected to this interior-state classification.
The implementation's aggregate indeterminate count also includes
these boundary candidates; it is therefore broader than the
number of interior representatives with indeterminate spectra.

If $d=0$, the compatibility class is a singleton with no
within-class perturbation directions. It cannot contain two
distinct equilibria and cannot yield a positive multistability
label.

The stability tolerance is a practical classification threshold,
not a rigorous eigenvalue-error bound. Together with residual
verification and finite-resolution clustering, the analysis
provides numerical evidence of local stability. It does not
determine attraction basins, certify global stability, or
establish switching or hysteresis behavior.

\end{document}